\documentclass[reprint,aps,prb,twocolumn,superscriptaddress,showpacs]{revtex4-1}
\usepackage{graphicx}
\graphicspath{{Figures/}}
\usepackage{dcolumn}
\usepackage{bm}
\usepackage{color}
\usepackage{amssymb} 
\usepackage{amsmath}
\usepackage{verbatim}
\usepackage{float}
\usepackage{subfig}
\usepackage{lmodern}
\usepackage[utf8]{inputenc}
\usepackage[T1]{fontenc}
\usepackage[english]{babel}

\usepackage{ragged2e}
\DeclareCaptionJustification{justified}{\justifying}
\newcommand{\la}{\langle}
\newcommand{\ra}{\rangle}
\newcommand{\rt}{\right}
\newcommand{\lf}{\left}
\newcommand{\Tr}{\operatorname{Tr}}

\renewcommand{\k}{\mathbf k}
\newcommand{\s}{\text{s}}
\newcommand{\ka}{\boldsymbol{\kappa}}
\newcommand{\ks}{{\boldsymbol{\kappa}_{\text{s}}}}
\renewcommand{\i}{\text{in}}

\newcommand{\br}{\mathbf r}
\newcommand{\el}{\text{el}}
\newcommand{\eq}[1]{Eq.~(\ref{#1})}
\newcommand{\Nel}{N_{\text{el}}}
\newcommand{\Hel}{\hat H_{\text{el}}}
\newcommand{\Hdel}{\hat H_{\text{el-em}}}
\newcommand{\Hint}{\hat H_{\text{int}}}
\newcommand{\Hem}{\hat H_{\text{em}}}
\newcommand{\Aem}{\hat{\mathbf A}_{\text{em}}}

\newcommand{\commentout}[1]{\ignorespaces}

\begin{document}
\title{Theory of x-ray scattering from laser-driven electronic systems}
\author{Daria Popova-Gorelova}
\email[]{daria.gorelova@desy.de}
\affiliation{Center for Free-Electron Laser Science, DESY, Notkestrasse 85, D-22607 Hamburg, Germany}
\affiliation{The Hamburg Centre for Ultrafast Imaging, University of Hamburg, Luruper Chaussee 149, D-22761 Hamburg, Germany}
\author{David A.~Reis}
\affiliation{PULSE Institute, SLAC National Accelerator Laboratory, Menlo Park, California 94025, USA}
\affiliation{Department of Applied Physics, Stanford University, Stanford, California 94305, USA}
\affiliation{Department of Photon Science, Stanford University, Stanford, California 94305, USA}
\author{Robin Santra}
\email[]{robin.santra@cfel.de}
\affiliation{Center for Free-Electron Laser Science, DESY, Notkestrasse 85, D-22607 Hamburg, Germany}
\affiliation{The Hamburg Centre for Ultrafast Imaging, University of Hamburg, Luruper Chaussee 149, D-22761 Hamburg, Germany}
\affiliation{Department of Physics, University of Hamburg, Jungiusstrasse 9, D-20355 Hamburg, Germany} 
\date{\today}

\begin{abstract}

We describe, within the framework of quantum electrodynamics, an interaction between a nonresonant hard x-ray pulse and an electronic system in the presence of a temporally periodic laser field driving electron dynamics in this system. We apply Floquet theory to describe the laser-driven electronic system, and then obtain the scattering probability of an arbitrary nonresonant x-ray pulse from such a system employing the density-matrix formalism. We show that the scattering probability can be connected to the time-dependent electron density of the driven electronic system only under certain conditions, in particular, if the bandwidth of the probe x-ray pulse is sufficiently narrow to spectroscopically resolve transitions to different final states. A special focus is laid on application of the theory to laser-driven crystals in a strongly nonperturbative regime. We show how the time-dependent electron density of a crystal can be reconstructed from energy-resolved scattering patterns. This is illustrated by a calculation of a diffraction signal from a driven MgO crystal. 
 
\end{abstract}
\maketitle

\section{Introduction}

An electronic system exposed to a periodic laser excitation is characterized by Floquet states, which can be seen as entangled states of electronic states and laser field photons \cite{ShirleyPR65, SantraPRA04}. Floquet states are a powerful theoretical concept to describe, on the one hand, quantum engineering of novel states of matter aided by a periodic excitation, and, on the other hand, nonperturbative processes driven by an intense laser field. The former class of processes includes, for example, creation of artificial magnetic fields and topological band structures \cite{OkaPRB09, StruckScience11, AidelsburgerPRL11, WangScience13, GoldmanPRX14, HuebenerNature17, OkaArxiv18}. The latter field of application of the Floquet theory is strong-field phenomena which cannot be understood by means of the conventional perturbation theory, such as high-order nonlinear optical processes in atomic, molecular and solid state systems \cite{FaisalPRA97, ChuPhRep04, SantraPRA04, ButhPRL07, SmirnovaNature09, GhimireNature11, SchubertNature14}.  

X-ray free-electron lasers, capable of producing pulses of hard x rays with angstrom wavelengths, offer unprecedented opportunities for imaging electronic structure of molecules and solids with atomic resolution \cite{CorkumNature07, GaffneyScience07, ChapmanNature06, VrakkingNature12, LeoneNature14, KowalewskiChRev17,NevillePRL18}. In this paper, we analyze an interaction of a nonresonant hard x-ray pulse with a system characterized by Floquet states in order to explore the opportunities to obtain temporal and spatial information about such a system.

The interaction between nonstationary electronic systems and x-ray pulses has already been analyzed in several studies \cite{DixitPNAS12, PopovaGorelovaPRB15_1, PopovaGorelovaPRB15_2}. However, these studies consider processes in which a pump pulse first brings an electronic system to a nonstationary state {\it triggering} its dynamics, and a probe x-ray pulse interacts with the system only after the action of the pump pulse. Here, we investigate a different process, in which pump and probe pulses act on a system simultaneously, whereby the pump pulse is a periodic {\it driving} force. Our analysis is performed within the framework of quantum electrodynamics (QED) and the density matrix formalism \cite{Mandel}, which have been demonstrated to be necessary for a correct description of the interaction of a nonstationary electronic system and x-ray pulses \cite{DixitPNAS12, Popova-GorelovaAppSci18}. In addition, it allows us to obtain expressions valid for arbitrary x-ray pulses, such as pulses of ultrashort time duration or having long coherence times, which are especially relevant for modeling of experiments at x-ray free electron lasers.

X-ray scattering from an electronic system interacting with an optical pulse in the linear regime was analyzed in the 1970s within a semiclassical theory \cite{FreundPRL70,EisenbergerPRA71}. This process was shown to lead to an x-ray and optical wave-mixing signal, which was connected to optically-induced charge densities. Our analysis in the present paper demonstrates that this connection is correct only under certain circumstances. An experiment, in which an x-ray pulse and an optical pulse simultaneously interacted with a crystal leading to x-ray and optical wave mixing has recently been realized at the x-ray-free electron laser facility Linac Coherent Light Source (LCLS) \cite{GloverNature12}. In this experiment, a linear effect of the optical field on the crystal manifested itself as a sum-frequency signal in x-ray diffraction. Our theory provides an interpretation of this experiment, as we will discuss in detail. 

Our study allows to describe x-ray diffraction from an electronic system not only in the regime of linear coupling to a driving field, but also in the high-order nonlinear interaction regime. The process of high harmonic generation (HHG), which serves, for example, for the generation of isolated attosecond pulses \cite{CorkumNature07,KrauszRMP09}, is essential for attosecond science and technology. Since the demonstration of HHG in bulk solids \cite{GhimireNature11}, it has attracted much attention, owing to its potential for producing attosecond pulses with a higher efficiency in comparison to that provided by gas-phase HHG. Furthermore, there is an intense discussion about the mechanism of HHG in solids \cite{SchubertNature14, VampaPRB15, McDonaldPRA15, NdabashimiyeNature16, TamayaPRL16, LuuPRB16, Tancogne-DejeanPRL17}. We have chosen a MgO crystal interacting with an infrared pulse in the regime of high-order harmonic generation \cite{GhimireJPhB14, YouNature16} to illustrate the possibilities of nonresonant x-ray scattering to probe electron dynamics in a crystal in such a regime. 

Although a special focus of this work is laid on laser-driven crystals, the expressions we derive are general for any electronic systems driven by a temporally periodic electromagnetic field. Therefore, our study can be applied to the development of techniques to image different types of driven systems such as atoms, molecules or quantum-engineered materials by means of nonresonant x-ray scattering. 

This article is organized as follows. In Section \ref{Section QEDFloquet}, we represent a driven electronic system within the Floquet formalism and the QED framework. We use this representation in Section \ref{Section Nonres} to describe the scattering probability of an arbitrary nonresonant x-ray pulse ({\it e.g.},~of arbitrary duration, coherence properties {\it etc.}) from such a system and consider some special cases following from this expression. In Section \ref{Section_Floquet-Bloch}, we apply our theory to describe nonresonant scattering from a laser-driven crystal. The particular case when the time-dependent electron density of a crystal can be reconstructed is described in Section \ref{Subsec_quasieel_crystal}. The results of Section \ref{Subsec_quasieel_crystal} are illustrated in Section \ref{Section_MgO} by a calculation of nonresonant x-ray scattering from a MgO crystal driven by an infrared pulse in a strongly nonlinear regime.

\section{Optically driven electronic system treated within the Floquet formalism in the QED framework}
\label{Section QEDFloquet}

Within the framework of quantum electrodynamics, the Hamiltonian describing the interaction of an electronic system with a single-mode electromagnetic field is
\begin{align}
&\Hdel = \Hel+\Hint+\Hem,\label{HamiltonianDrivenSystem}\\
&\Hem = \omega\hat a_{\ka_0,s_0}^\dagger\hat a_{\ka_0,s_0},\\
&\Hint = \overline\alpha\int d^3r\hat \psi^\dagger(\mathbf r)\lf(\Aem(\mathbf r)\cdot\mathbf p\rt)\hat \psi(\mathbf r).
\end{align}
Here, $\Hel$ is the Hamiltonian of the electronic system, $\Hem$ is the Hamiltonian of the electromagnetic field, and $\Hint$ describes the interaction between the electromagnetic field and the electronic system. At this stage, we do not apply the dipole approximation to the Hamiltonian $\Hint$, which is not a necessary condition to treat the problem within the Floquet formalism. $\hat a_{\ka,s}^\dagger$ ($\hat a_{\ka,s}$) creates (annihilates) a photon with wave vector $\ka$ and polarization $s$. We assume that only the $\ka_0$, $s_0$ mode with a corresponding polarization vector $\boldsymbol\epsilon_0$ and the energy $\omega = |\ka_0|c$, where $c$ is the speed of light, is occupied in the driving electromagnetic field, and that the state of the field is described by a single-mode coherent state $|\alpha,t\ra$. $\Aem (\mathbf r)$ is the vector potential operator of the electromagnetic field, $\mathbf p$ is the canonical momentum of an electron, $\hat \psi^\dagger$ ($\hat \psi$) is the electron creation (annihilation) field operator, and $\overline\alpha$ is the fine-structure constant. We neglect the $\Aem^2$ contribution for the optical field. We use atomic units for this and the following expressions. 

The Hamiltonian $\Hdel$ can be represented as a matrix in the basis $|\Phi_{n}\ra|N-\mu\ra$, which are product states formed by many-body eigenstates of $\Hel$, $|\Phi_{n}\ra$, and Fock states of the mode $\ka_0$, $s_0$, $|N-\mu\ra$, where $\mu$ is an integer \cite{SantraPRA04, ShirleyPR65}. $N$ is an integer approximating the average number of photons, $\la \alpha,t|\hat a_{\ka_0,s_0}^\dagger\hat a_{\ka_0,s_0}|\alpha,t\ra$, in the mode $\ka_0$, $s_0$.  For ultrafast dressing experiments that are typically carried out with light pulses with energies of the order of 1 mJ and at photon energies of around 1 eV,  $N$ is of the order of $10^{15}$. $|\mu|$ is related to the number of photons involved in the interaction between the electronic system and the electromagnetic field, which, in practice, is limited by some value resulting in $-\mu_{\text{max}}\le\mu\le\mu_{\text{max}}$. It must be satisfied that $N\gg\mu_{\text{max}}$ for the coupling elements of $\Hdel$ to be independent of $\mu$. With these conditions, $\Hdel$ is a block matrix with a quite sparse structure

\begin{widetext}
\begin{align}
&\left(
\begin{array}{ccccccc}
\ddots & \ddots &\ddots& \ddots& \ddots& \ddots & \ddots\\
\ddots & \mathbf E+(N-2)\omega\mathbf I & \mathbf T & \boldsymbol 0 & \boldsymbol 0 & \boldsymbol 0 & \ddots \\
\ddots &\mathbf T^\dagger & \mathbf E +(N-1) \omega\mathbf I & \mathbf T & \boldsymbol 0 & \boldsymbol 0 &  \ddots \\
\ddots &\boldsymbol 0 & \mathbf T^\dagger & \mathbf E+N\omega\mathbf I & \mathbf T & \boldsymbol 0 &  \ddots \\
\ddots &\boldsymbol 0 & \boldsymbol 0&\mathbf T^\dagger & \mathbf E+(N+1)\omega\mathbf I & \mathbf T  & \ddots \\
\ddots &\boldsymbol 0 & \boldsymbol 0 & \boldsymbol 0 &\mathbf T^\dagger & \mathbf E+(N+2)\omega\mathbf I & \ddots \\
\ddots  & \ddots &\ddots& \ddots& \ddots& \ddots &\ddots
\end{array}
\right),
\label{Eq_FloquetMatrix}
\end{align}
\end{widetext}
where $\mathbf E$ is a diagonal matrix with the diagonal elements being the eigenenergies $E_{\Phi_n}$ of the Hamiltonian $\Hel$ of the electronic system with $\Nel$ electrons, $\mathbf I$ is a unit matrix and $\boldsymbol 0$ is a zero matrix. $\mathbf T$ is a matrix with elements $t_{n',n} =\la N-\mu-1| \la \Phi_{n'}| \Hint |\Phi_n\ra|N-\mu\ra \propto \sqrt N \int d^3 r\la \Phi_{n'}|e^{i\ka_0\cdot\br}\hat\psi^\dagger(\br) (\boldsymbol\epsilon_0\cdot\mathbf p)\hat\psi(\br)|\Phi_n\ra$, where $t_{n,n}$ can be nonzero beyond the dipole approximation. Due to the approximations mentioned above, the interaction of the electromagnetic field with the electronic system is treated in the classical limit of the QED. The semiclassical treatment within the Floquet formalism is indeed a very good approximation to describe the interaction of atoms, molecules and solid-state systems with strong fields. This includes the regime of high harmonic generation as discussed, for instance, in Refs.~\onlinecite{ChuPhRep04, SpringerBookAMO, FaisalPRA97, HsuPRB06}.

The eigenstates of $\Hdel$ are Floquet states represented as a superposition
\begin{align}
|\Psi_K\ra = \sum_{n,\mu}C^K_{ n,\mu}|\Phi_{n}\ra|N-\mu \ra\label{Eq_PsiK_gen}.
\end{align}
Due to the periodic structure of the matrix in \eq{Eq_FloquetMatrix}, each Floquet eigenstate $\Psi_{K}$ has replica states, which are physically equivalent to each other. If $\Psi_{K_0}$ is some reference eigenstate with energy $E_{K_0}$, then its replicas are 
\begin{align}
|\Psi_{K_{\Delta\mu}}\ra &= \sum_{n,\mu}C^{K_{\Delta\mu}}_{ n,\mu}|\Phi_{n}\ra|N-\mu \ra\nonumber\\
 &= \sum_{n,\mu}C^{K_0}_{ n,\mu+\Delta\mu}|\Phi_{n}\ra|N-\mu \ra\label{EqReplicaStates}
\end{align}
with the corresponding eigenenergies $E_{K_{\Delta\mu}} = E_{K_0}+\Delta\mu\omega$, where $\Delta\mu$ is an integer. 

Let us now determine the state of the light-driven electronic system $|\Psi_0,t\ra$, which is the solution of the time-dependent Schr\"odinger equation $i\partial |\Psi_0,t\ra/\partial t = \Hdel|\Psi_0,t\ra$, under the assumption that the state of the electronic system at time $t=0$ is known. We consider a general case, then this state of the electronic system is a superposition of its electronic eigenstates $|\Psi_{\text{el}},0\ra = \sum_n\widetilde C_n|\Phi_n\ra$. It is assumed that the state of the electromagnetic field $|\alpha,t\ra$, which can be represented as $\sum_{\mu}A_{N-\mu}e^{-i(N-\mu)\omega t}|N-\mu\ra$, is unaffected by the interaction with the electronic system. Thus, $|\Psi_0,t\ra$ is given by
\begin{align}
|\Psi_0,t\ra = |\Psi_{\text{el}},t\ra|\alpha,t\ra,\label{Eq_Psi0Factorized}
\end{align}
which can be applied to determine the boundary condition $|\Psi_0,0\ra$ for the time-dependent Schr\"odinger equation. The approximation that $\sum_{\mu}A^*_{N-\mu}A_{N-\mu+\Delta\mu}\approx 1$ independently of $\Delta\mu$ for very large $N$ and $|\Delta\mu|\ll N$ leads to the solution \cite{ShirleyPR65}
\begin{align}
|\Psi_0,t\ra = \sum_{K_0}\mathcal C_{K_0}e^{-iE_{K_0} t}|\Theta_{K_0},t\ra.\label{Eq_wavepacket}
\end{align}
Here, the state of the light-driven electronic system $|\Psi_0,t\ra$ is represented as a superposition of Fourier series 
\begin{align}
|\Theta_{K_0},t\ra = \sum_{\mu,n} C_{n,\mu}^{K_0} e^{-i\mu\omega t}|\Phi_n\ra|\alpha,t\ra,\label{Eq_Theta}
\end{align}
which involve physically equivalent Floquet states, with expansion coefficients
\begin{align}
\mathcal C_{K_0}= \sum_{\mu,n}\widetilde C_nC^{K_{0}*}_{n,\mu},
\end{align}
which are determined by the state of the electronic system at time $t=0$. The state $|\Psi_0,t\ra$ does not depend on the choice of a reference state $K_0$ among its replicas.


The time-dependent electron density of the light-driven system is given by $\rho(\mathbf r,t) = \la \Psi_0,t|\hat\psi^\dagger(\br)\hat\psi(\br)|\Psi_0,t\ra$. We obtain in Appendix \ref{App_density0} that 
\begin{align}
&\rho(\br,t) = \sum_{K_0,I_0}\mathcal C_{K_0}^*\mathcal C_{I_0}e^{i(E_{K_0}-E_{I_0})t}\rho_{K_0I_0}(\br,t),\label{EqDensity}
\end{align}
where $\rho_{K_0I_0}(\br,t)$ can be represented as a Fourier series
\begin{align}
&\rho_{K_0I_0}(\br,t) = \sum_{\Delta\mu}e^{i\Delta\mu\omega t}\widetilde\rho_{K_0I_0}(\br,\Delta\mu)\label{EqDensityKI}
\end{align}
with amplitudes
\begin{align}
&\widetilde\rho_{K_0I_0}(\br,\Delta\mu) = \sum_{n,n',\mu}C^{K_0*}_{n',\mu+\Delta\mu}C^{I_0}_{n,\mu}\la \Phi_{n'}|\hat\psi^\dagger(\br)\hat\psi(\br)|\Phi_n\ra.\label{EqDensityFourierMB}
\end{align}

\section{Nonresonant x-ray scattering from a laser-driven electronic system}
\label{Section Nonres}

If the driven electronic system is probed by means of high-energy nonresonant x-ray scattering, then the total Hamiltonian of the whole system, matter and light, is given by
\begin{align}
&\hat H = \Hdel+\hat H_{\text{int}x}+\hat H_{x},\\
&\hat H_{\text{int}x}=\frac{\overline\alpha^2}{2}\int d^3 r\hat \psi^\dagger(\mathbf r)\hat{\mathbf A}_x^2(\mathbf r)\hat \psi(\mathbf r),\\
&\hat H_{x} = \sum_{\ka_x,s}\omega_{\ka_x}\hat a_{\ka_x,s}^\dagger\hat a_{\ka_x,s}.
\end{align}
Here, $\hat H_{x}$ is the Hamiltonian of the x-ray field, $\hat H_{\text{int}x}$ is the interaction Hamiltonian between the electronic system and the x-ray field in a high-energy nonresonant regime, and $\hat{\mathbf A}_x$ is the vector potential of the x-ray field.

We derive the probability of observing a scattered photon with momentum $\ka_{\s}$, $P(\ka_{\s})$, within the density-matrix formalism \cite{Mandel} as
\begin{align}
&P(\ks) = \sum_{s_\s,\{n'_x\},F}\la\Psi_F;\{n'_x\}|\hat\rho_1|\Psi_F;\{n'_x\}\ra,\label{Prob_denmatrix}
\end{align}
where $\{n'_x\}$ is the x-ray field configuration that has one photon in the scattering mode $\ka_{\s}$ and the sum is over all possible final states $\Psi_F$, which are the eigenstates of the Hamiltonian $\Hdel$. 
\begin{align}
&\hat\rho_1 =\lim_{t_f\rightarrow\infty}\sum_{\{ n_x \}, \{\widetilde n_x \}}\rho^x_{\{ n_x \}, \{\widetilde n_x \}}|\Psi_{\{n_x\}}^{(1)},t_f\ra\la\Psi_{\{\widetilde n_x\}}^{(1)},t_f|,\label{rho1_Eq}
\end{align}
is the total density matrix of the driven electronic system and the x-ray field, which is evaluated within the first-order time-dependent perturbation theory using the interaction Hamiltonian with the x-ray field $\hat H_{\text{int}x}$ as the perturbation. $\{ n_x \}$ and $\{\widetilde n_x \}$ are sets of Fock states that specify the number of photons in all initially occupied modes of the x-ray field with a distribution $\rho^x_{\{ n_x \}, \{\widetilde n_x \}}$, and  
\begin{align}
&|\Psi_{\{n_x\}}^{(1)},t_f\ra = -i\int_{-\infty}^{t_f} dte^{i(\Hdel+\hat H_{x})(t-t_f)}\hat H_{\text{int}x}\label{Psi1_Eq}\\
&\qquad\qquad\qquad\qquad\times e^{-i(\Hdel+\hat H_{x})t}|\Psi_0,t\ra|\{n_x\}\ra\nonumber.
\end{align} 

The formalism to describe the scattering probability $P(\ka_{\s})$ is similar to the one applied in Ref.~\cite{DixitPNAS12}, where the interaction of a nonstationary electronic system with a nonresonant x-ray pulse has also been considered. However, we analyze a regime where a nonstationary electronic system interacts with a probe and a pump pulse simultaneously in contrast to Ref.~\cite{DixitPNAS12}, where it is assumed that the probe pulse arrives after the pump pulse. In Appendix \ref{App_scatt_prob}, we derive a general expression for the scattering probability of a probe nonresonant hard-x-ray pulse of arbitrary coherent properties and duration, which is applicable for both time-resolved and -unresolved measurements, 
\begin{widetext}
\begin{align}
&P(\ks)
=P_0\sum_{F_0}\int_{-\infty}^{+\infty} dt_1\int_{-\infty}^{+\infty} dt_2\int d^3 r_1\int d^3 r_2  
G^{(1)}(\br_2,t_2,\br_1,t_1)e^{i\omega_\ks (t_1-t_2)-i\ks\cdot(\br_1-\br_2)} M_{F_0}^*(\br_2,t_2) M_{F_0}(\br_1,t_1),\label{Prob_General}
\end{align}
\end{widetext}
where $P_0 = \sum_{s_\s}|(\boldsymbol\epsilon_{x\i}\cdot \boldsymbol\epsilon^*_{x\ks,s_\s})|^2\omega_\ks^2/(4\pi^2\omega_{x\i}^2c^3)$, $\boldsymbol\epsilon_{x\i}$ is the mean polarization vector of the incoming x-ray beam, the sum over $s_\s$ refers to the sum over polarization vectors of the scattered photons $\boldsymbol\epsilon^*_{x s_\s}$, $\omega_{x\i}$ is the mean photon energy of the incoming x-ray beam and $\omega_{\ks}$ is the energy of the scattered photon. The summation is over such final Floquet states $F_0$ that their replica states $F_{\Delta\mu\neq0}$ do not enter the summation. The scattering probability does not depend on the choice of the reference state $F_0$ among its replica states.


$G^{(1)}(\br_2,t_2,\br_1,t_1)$ is the first-order x-ray field correlation function \cite{GlauberPhRev63, Loudon}.  It depends on the probe-pulse arrival time $t_p$ and provides the dependence of the scattering probability on $t_p$ in the case of a time-resolved measurement. The function $M_{F_0I_0}(\br,t)$ analogously to the electron density in Eqs.~(\ref{EqDensity}) and (\ref{EqDensityFourierMB}) can be represented as a sum of Fourier series
\begin{align}
M_{F_0}(\br,t)=& \sum_{I_0}\mathcal C_{I_0}e^{i(E_{F_0}-E_{I_0}) t}\sum_{\Delta\mu}e^{i\Delta\mu\omega t} \widetilde M_{F_0I_0}(\mathbf r,\Delta\mu)\label{Eq_M_FI}
\end{align}
with amplitudes 
\begin{align}
\widetilde M_{F_0I_0}(\mathbf r,\Delta\mu) =   \sum_{n,n',\mu}C^{F_0^*}_{n',\mu+\Delta\mu}C^{I_0}_{n,\mu}\la \Phi_{n'}|\hat\psi^\dagger(\br)\hat\psi(\br)|\Phi_n\ra. 
\end{align}
The electron density is related to these functions via $\rho(\br,t) = \sum_{F_0}\mathcal C_{F_0}^*M_{F_0}(\br,t)$ [{\it cf.~}\eq{EqDensity}]. However, the x-ray scattering probability in \eq{Prob_General} is in general not connected to the electron density, because, first, the coefficients $\mathcal C_{F_0}^*$ do not enter this equation, and, second, the summation over $F_0$ is incoherent. In other words, the time-dependent electron density is not the quantity that determines the scattering probability signal, which can be different at equal electron densities \cite{DixitPNAS12, PopovaGorelovaPRB15_1}.

\subsection{Perfectly coherent x-ray probe pulse}

Let us consider a perfectly coherent x-ray probe pulse, which results in the factorizable correlation function
\begin{align}
G^{(1)}(\br_2,t_2,\br_1,t_1) 
= &\frac{\mathcal E^*_{x}(\mathbf r_2,t_2-t_p)}2 e^{i\omega_{x\i}t_2-i\boldsymbol\kappa_{\i}\cdot\br_2}\\
&\times 
\frac{\mathcal E_{x}(\mathbf r_1,t_1-t_p)}2 e^{-i\omega_{x\i}t_1+i\boldsymbol\kappa_{\i}\cdot\br_1},\nonumber
\end{align}
where $\mathcal E_x(\mathbf r,t-t_p)$ is the amplitude of the x-ray field, which does not noticeably vary in comparison to the size of the object positioned at $\br_0$. In this case, the probability is given by
\begin{align}
&P(\mathbf q_x)
=P_0\sum_{F_0}\lf| \sum_{I_0,\Delta\mu}  \mathcal M_{F_0I_0}(\mathbf q_x,\Delta\mu)\,\widetilde{\mathcal E}_x (\Omega_{F_0I_0}+\Delta\mu\omega)   \rt|^2\label{Prob_General_coherent},
\end{align}
where $\Omega_{F_0I_0} = \omega_\ks-\omega_{x\i}+E_{F_0}-E_{I_0}$, $\mathbf q_x = \ka_\i-\ks$ and 
\begin{align}
\mathcal M_{F_0I_0}(\mathbf q_x,\Delta\mu) =\mathcal C_{I_0} \int d^3 re^{i \mathbf q_x\cdot\br}  \widetilde M_{F_0I_0}(\mathbf r,\Delta\mu). 
\end{align}
Here, $\widetilde{\mathcal E}_x(\omega')$ is the Fourier transform of the electric-field amplitude of the x-ray field: 
\begin{align}
\widetilde{\mathcal E}_x(\Omega_{F_0I_0}+\Delta\mu\omega)=\int_{-\infty}^{\infty}dt \mathcal E_x(\mathbf r_0,t-t_p) e^{i(\Omega_{F_0I_0}+\Delta\mu\omega)t}.
\end{align}
In the case of a time-resolved measurement, this function provides the dependence of the scattering probability on the probe-pulse arrival time $t_p$.

\begin{figure}[t]
\centering
\includegraphics[width=0.5\textwidth]{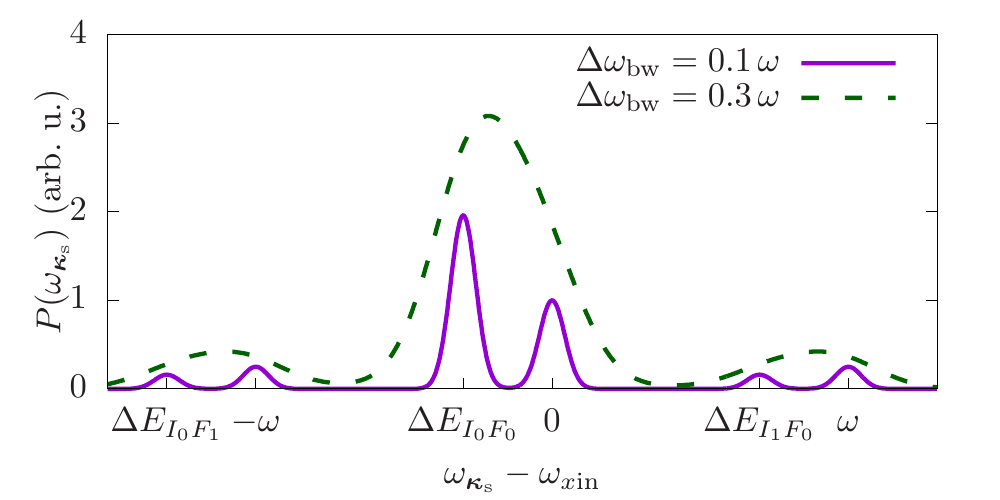}
\caption{Illustration of the scattering probability as a function of $\omega_\ks-\omega_{x\i}$ for two different probe-pulse bandwidths $\Delta\omega_{\text{bw}}$. $\Delta E_{I_{\Delta\mu'}F_{\Delta\mu}} = E_{I_0}-E_{F_0}+(\Delta\mu'-\Delta\mu)\omega$.}
\label{Fig_peaks}
\end{figure}

Let us illustrate the dependence of the scattering probability in \eq{Prob_General_coherent} on the scattering energy $\omega_\ks$ using the following example. Let us assume that the Hamiltonian of a laser-driven system has just two types of eigenstates $I_{\Delta\mu}$ and $F_{\Delta\mu'}$, where the minimum energy splitting among these states is $E_{F_0}-E_{I_0}=0.3\omega$. We assume that only one nonzero coefficient, $\mathcal C_{I_0} = 1$, enters $|\Psi_0,t\ra$, and, thus, the state of the system is described by a single series $|\Theta_{I_0},t\ra$ comprising states $I_{\Delta\mu}$ [{\it cf.}~\eq{Eq_wavepacket}]. The solid violet curve in Fig.~\ref{Fig_peaks} shows $P(\omega_\ks)$ at a fixed scattering angle assuming a Gaussian-shaped probe pulse with a bandwidth (the full width at half maximum of the intensity) $\Delta\omega_{\text{bw}} = 0.1\omega$. The values of the functions $\mathcal M_{I_0I_0}(\mathbf q_x,\Delta\mu)$ and $\mathcal M_{F_0I_0}(\mathbf q_x,\Delta\mu)$ are chosen randomly and their variation as a function of $\omega_\ks-\omega_{x\i}$ within the probe-pulse bandwidth is assumed to be negligible. As shown in the plot, the scattering probability consists of a series of peaks centered at $\Delta\mu\omega$ and $E_{I_0}-E_{F_0}+\Delta\mu\omega$. The width of these peaks is equal to the bandwidth of the probe pulse, and their amplitudes are time-independent and proportional to $|\mathcal M_{I_0I_0}(\mathbf q_x,\Delta\mu)|^2$ and $|\mathcal M_{F_0I_0}(\mathbf q_x,\Delta\mu)|^2$, respectively. Figure~\ref{Fig_peaks} shows the peaks corresponding to $\Delta\mu=-1$, 0 and 1.

The dashed curve in green in Fig.~\ref{Fig_peaks} shows the scattering probability assuming a probe pulse with a bandwidth $\Delta\omega_{\text{bw}} = 0.3\omega$, all other parameters being the same as for the solid violet curve. In contrast to the previous example, the contributions to the scattering probability due to transitions with different final states intermix in the spectrum and cannot be separated. This illustrates that if the bandwidth of the probe x-ray pulse is not considerably smaller than the difference $|E_{F_{\Delta\mu}}-E_{F'_{\Delta\mu'}}|$ between energies of final states $F_{\Delta\mu}$ and $F'_{\Delta\mu'}$ for any $\Delta\mu$ and $\Delta\mu'$, it is not possible to spectroscopically distinguish between the contributions to the scattering probability due to transitions to final states $F_{\Delta\mu}$ and due to transitions to final states $F'_{\Delta\mu'}$. At the same time, the scattering signal is time-resolved unlike the previous case, since the amplitudes of the peaks on the dashed curve in green depend on the probe-pulse arrival time $t_p$ via the interference terms proportional to $\widetilde{\mathcal E}^*_x(\Omega_{F_0I_0}+\Delta\mu'\omega)\widetilde{\mathcal E}_x(\Omega_{I_0I_0}+\Delta\mu\omega)$.

\subsection{Quasielastic scattering by a narrow-bandwidth probe x-ray pulse}
Let us now consider a probe x-ray pulse with a bandwidth smaller than $\omega$, energy differences between any states $I_{\Delta\mu}$ and $K_{\Delta\mu'}$ comprising the wave packet $\Psi_0$, and energy differences between these states and other Floquet eigenstates $F_{\Delta\mu''}$ for any $\Delta\mu$, $\Delta\mu'$ and $\Delta\mu''$, so that one can spectroscopically distinguish between scattering to different final states. In this case, it is possible to define the probability of quasielastic scattering $P_{qe}$ as the probability to separately measure scattering events with final states being the ones comprising the wave packet $\Psi_0$. Thus, the probability of quasielastic scattering is obtained by replacing the summation over final states $F_0$ in the expression for the total scattering probability in \eq{Prob_General} for the summation over states $K_0$ for which $\mathcal C_{K_0}\neq0$. 

The probability of quasielastic scattering is given by a sum of terms, which include integrals $\int dt_1 \int d t_2G^{(1)}(\br_2,t_2,\br_1,t_1) $ $\widetilde M_{K_0I_0}^*(\br_2,\Delta\mu')\widetilde M_{K_0I_0}(\br_1,\Delta\mu)e^{i(\Delta\mu t_1-\Delta\mu' t_2)\omega t}$ [{\it cf.}~\eq{Eq_M_FI}]. Due to the condition that the probe-pulse bandwidth is smaller than energy differences between any states $I_{\Delta\mu}$ and $K_{\Delta\mu'}$, these integrals are nonzero only for $\Delta\mu=\Delta\mu'$ and $K_0=I_0$, and the interference terms disappear. Thus, taking into account that $\widetilde M_{I_0I_0}=\widetilde\rho_{I_0I_0}$, the probability of quasielastic scattering is given by
\begin{widetext}
\begin{align}
P_{qe}(\ks)
=P_0\int_{-\infty}^{+\infty} dt_1\int_{-\infty}^{+\infty} dt_2\int d^3 r_1\int d^3 r_2 & 
G^{(1)}(\br_2,t_2,\br_1,t_1)\sum_{\Delta\mu}e^{i(\omega_\ks+\Delta\mu) (t_1-t_2)-i\ks\cdot(\br_1-\br_2)}\nonumber\\
&\times\sum_{I_0}|\mathcal C_{I_0}|^2\widetilde \rho_{I_0I_0}^*(\br_2,\Delta\mu)\widetilde \rho_{I_0I_0}(\br_1,\Delta\mu).\label{Prob_qe_General}
\end{align}
\end{widetext}
It is still not connected to the electron density in \eq{EqDensity} unless $\mathcal C_{I_0} = \delta_{I_0,K_0}$ for some state $K_0$. Thus, if the state of a light-dressed electronic system is described by a superposition of physically inequivalent Floquet states, the scattering signal from it cannot be related to its electronic density. 

\subsection{Quasielastic scattering from a light-dressed electronic system described by a single family of Floquet states}

But let us now consider a light-dressed electronic system in a state $|\Psi_0,t\ra$ described by a single series $|\Theta_{I_0},t\ra$ of physically equivalent Floquet states meaning that $\mathcal C_{I_0} = \delta_{I_0,K_0}$ for some state $K_0$. In this situation, the wave function of the electronic system evolves in time periodically with the frequency $\omega$ and is given by a superposition of electronic eigenstates with time-dependent coefficients $|\Psi_{\el},t \ra= \sum_{\mu,n}C^{I_0}_{n,\mu}e^{-i\mu\omega t}|\Phi_n\ra$ [{\it cf.~}Eqs.~(\ref{Eq_Psi0Factorized})-(\ref{Eq_Theta})]. In particular, its time-dependent electronic density evolves periodically with the frequency $\omega$ and can be represented by a Fourier series $\rho(\br,t) = \sum_{\Delta\mu}e^{i\Delta\mu\omega t}\widetilde\rho(\br,\Delta\mu)$, where $\widetilde\rho(\br,\Delta\mu)=\widetilde\rho_{I_0I_0}(\br,\Delta\mu)$. Then, if the probe x-ray pulse has a bandwidth sufficiently narrow to separate the contribution due to quasielastic scattering, the probability of  quasielastic scattering,
\begin{widetext}
\begin{align}
P_{qe}(\mathbf q_x)
&=P_0\int_{-\infty}^{+\infty} dt_1\int_{-\infty}^{+\infty} dt_2\int d^3 r_1\int d^3 r_2 
 G^{(1)}(\br_2,t_2,\br_1,t_1)e^{i\omega_\ks (t_1-t_2)-i\ks\cdot(\br_1-\br_2)}
\rho(\br_1,t_1)\rho(\br_2,t_2),
\end{align}
\end{widetext}
does depend on the time-dependent electron density $\rho(\br,t)$. 

In addition, if the probe x-ray pulse is spatially uniform and perfectly coherent, then $P_{qe}$ is given by a series of peaks with a width equal to the bandwidth of the probe pulse centered at scattering energies $\omega_{x\i}+\Delta\mu\omega$
\begin{align}
P_{qe}(\mathbf q_x)
=&P_0 \sum_{\Delta\mu}|\widetilde{\mathcal E}_x (\omega_\ks-\omega_{x\i}+\Delta\mu\omega)|^2 \widetilde P_{qe}(\mathbf q_x,\Delta\mu),\label{Probab_via_el_den}
\end{align}
where the amplitudes of the peaks
\begin{align}
&\widetilde P_{qe}(\mathbf q_x,\Delta\mu) \propto\Biggl|  \int d^3 re^{i \mathbf q_x\cdot\br}\widetilde\rho(\br,\Delta\mu) \Biggr|^2\nonumber.
\end{align} 
are connected to the corresponding $\Delta\mu$-th amplitudes of the electron density, $\widetilde\rho(\br,\Delta\mu)$. Thus, in this situation, the probability of quasielastic scattering $P_{qe}$ provides the spatial and temporal Fourier transform of the time-dependent electron density of the driven electronic system evolving with the frequency $\omega$.

\subsection{Discussion}

To sum up, we considered in this Section a pump-probe experiment, in which a temporally periodic pump pulse drives electron dynamics in a system bringing it to a state $|\Psi_0,t\ra$, which is a superposition of Fourier series comprising physically equivalent Floquet states [{\it cf.~}Eq.~(\ref{Eq_wavepacket})]. A nonresonant hard x-ray pulse is used as a probe of the dynamics of the laser-driven electronic system. It induces transitions from the initial states $K_{\Delta\mu}$ to final states, which are either one of the eigenstates $K_{\Delta\mu'}$ comprising the wave packet $|\Psi_0,t\ra$ or to final Floquet states $F_{\Delta\mu''}$, which are different from any $K_{\Delta\mu'}$ states. We refer to the former events as quasielastic scattering. The contribution due to quasielastic scattering can be isolated from the total scattering signal only if the bandwidth of the probe pulse is considerably smaller than any energy splittings between $K_{\Delta\mu}$ and $F_{\Delta\mu'}$ states for any $\Delta\mu$ and $\Delta\mu'$. 

Only if the state of the driven electronic system is prepared in such a way that $|\Psi_0,t\ra$ may be expanded in terms of a single family of replica states, the probability of quasielastic scattering is connected to the time-dependent electron density. In the general case, which particularly applies to an ultrashort probe x-ray pulse, it may not be possible to spectroscopically distinguish between inelastic and quasielastic contributions to the scattering probability. Then, the scattering probability is determined by unseparable contributions determined by the functions $M_{F_0I_0}(\br,t)$ and cannot be connected to the time-dependent electron density. 

If the state of the driven electronic system $|\Psi_0,t\ra$ is a superposition of more than one Fourier series involving physically inequivalent Floquet states, neither the total nor the quasielastic scattering probability is connected to the electronic density. Then, it may not be advantageous to probe electron dynamics by separating quasielastic and inelastic contributions to the scattering probability. One may have to search for alternative ways to extract information about electron dynamics from a scattering signal \cite{Popova-GorelovaAppSci18}.


So far, we have not applied any assumptions concerning the electronic system. The expressions describing the interaction between a driven electronic system and a probe nonresonant x-ray pulse derived in this Section are general for any electronic system. In the next Section, we consider the particular case of a spatially periodic electronic system.

\section{Application to a spatially periodic electronic system}
\label{Section_Floquet-Bloch}

Let us specifically consider the case when the driven electronic system is a crystal described by the effective one-electron Hamiltonian
\begin{align}
&\Hel = \int d^3 r \hat\psi^\dagger(\br)[\mathbf p^2/2+V_c(\br)]\hat\psi(\br),
\end{align}
where $V_c(\br) = V_c(\br+\mathbf R)$ is a space-periodic crystal field potential, $\mathbf R$ is a lattice vector. We diagonalize the Hamiltonian in Eq.~(\ref{HamiltonianDrivenSystem}) using the basis set
\begin{align}
|\varphi_{m,\k,\mu }\ra = |\varphi_{m\k}\ra|N-\mu \ra,\label{Eq_basis_set_onebody}
\end{align}
where $|\varphi_{m\k}\ra$ are one-body eigenstates of the field-free Hamiltonian $\Hel$ such that $\k$ is the Bloch wave vector and $m$ is the band and spin index. According to the Bloch theorem \cite{Kittel}, the corresponding one-body wave function of $|\varphi_{m\k}\ra$ has the form $\varphi_{m\k}(\br)=e^{i\k\cdot\mathbf r}u_{m\k}(\br)$, where $u_{m\k}(\br) = u_{m\k}(\br+\mathbf R)$ is a space-periodic function. 

The matrix elements of the Hamiltonian $\Hdel$ obtained within the QED picture are equivalent to those derived in Refs.~\onlinecite{HsuPRB06, FaisalPRA97, TzoarPRB75} within a semiclassical theory, for $N\gg 1$ and $|\mu|\ll N$, 
\begin{align}
&\la \varphi_{m,\k,\mu }| \Hel+\Hem | \varphi_{m,\k,\mu }\ra = E_{m,\k}+(N-\mu)\omega,\label{Eq_spatial_periodic_Ham1}\\
&\la\varphi_{m',\k,\mu+1 }| \Hint | \varphi_{m,\k,\mu }\ra \nonumber\\
&  =\sqrt{\frac{2\pi I_{\text{em}}}{\omega^2c}}\Big(\boldsymbol\epsilon_0\cdot\big[\k \delta_{m'm}+\mathbf D_{m'm}\big]\Big),\label{Eq_spatial_periodic_Ham2}\\
&\la\varphi_{m',\k,\mu }| \Hint | \varphi_{m,\k,\mu+1 }\ra\nonumber\\
& = \sqrt{\frac{2\pi I_{\text{em}}}{\omega^2c}}\Big(\boldsymbol\epsilon_0^*\cdot\big[\k \delta_{m'm}+\mathbf D_{m'm}\big]\Big).\label{Eq_spatial_periodic_Ham3}
\end{align}
Here, $\mathbf D_{m'm}= -i N_{\text{cells}}\int\limits_{V_{\text{cell}}} d^3 r u_{m',\k}^\dagger(\br) \boldsymbol\nabla u_{m,\k}(\br)$, where the integration is over the volume of the crystal unit cell, $V_{\text{cell}}$, and $N_{\text{cells}}$ is the number of unit cells interacting with the driving electromagnetic field. Here, the interaction between the crystal and the driving electromagnetic field is described within the dipole approximation. We took into account that $I_{\text{em}}=N\omega c/V$, which is the intensity measured in units of $E_h/(t_{\text{au}}a_{\text{au}}^2) = 6.43641\times10^{15}$ W/cm$^2$ ($E_h$ is the Hartree energy, $t_{\text{au}}$ is the atomic unit of time and $a_{\text{au}}$ is the Bohr radius). Other matrix elements of $\Hdel$ are zero. Thus, one-body eigenstates of the Hamiltonian $\Hdel$ are
\begin{align}
|\phi_{i,\k}\ra = \sum_{m,\mu}c^i_{ m,\k,\mu}|\varphi_{m\k}\ra|N-\mu \ra\label{Eq_expansion_ob}
\end{align}
with corresponding eigenenergies $\varepsilon_{i,\k}$. The coefficients $c^i_{m,\k,\mu}$ are the solutions of the equation
\begin{align}
\sum_{m'\mu'}\la \varphi_{m,\k,\mu}|\Hdel|\varphi_{m',\k,\mu'}\ra c^i_{ m',\k',\mu'} = \varepsilon_{i,\k}c^i_{ m,\k,\mu}.
\end{align}
This model describes laser-induced electron dynamics only due to interband transitions vertical in the $\k$ space.

We prove in Appendix \ref{App_MF0I0} that 
\begin{align}
\widetilde\rho_{I_0I_0}(\br,\Delta\mu) 
&=  \sum_{i,\k} \sum_{m,m',\mu}c^{i*}_{m',\k,\mu+\Delta\mu}c^i_{m,\k,\mu}\label{rhoI0I0viaOneBody}\\
&\qquad\qquad\times u^\dagger_{m'\k}(\br)u_{m\k}(\br),\nonumber
\end{align}
where the summation is over such $i$ and $\k$ that the state $|\phi_{i,\k}\ra$ is occupied in $|\Psi_{I_0}\ra$, and
\begin{align}
\widetilde\rho_{K_0I_0\neq K_0}(\br,\Delta\mu) 
=& \sum_{m',m,\mu}c^{k*}_{m',\k',\mu+\Delta\mu}c^i_{m,\k,\mu}\label{rhoK0I0viaOneBody}\\
&\qquad\times u^\dagger_{m'\k'}(\br)u_{m\k}(\br)e^{i(\k-\k')\cdot\br},\nonumber
\end{align}
which is nonzero only if all the same one-body Floquet states are occupied in $|\Psi_{I_0}\ra$ and $|\Psi_{K_{\Delta\mu}}\ra$ except that the state $|\phi_{i,\k}\ra$ is occupied in $|\Psi_{I_0}\ra$ and not occupied in $|\Psi_{K_{\Delta\mu}}\ra$, and the state $|\phi_{k,\k'}\ra$ is occupied in $|\Psi_{K_{\Delta\mu}}\ra$ and not occupied in $|\Psi_{I_0}\ra$. 


\subsection{Total scattering probability of a coherent probe x-ray pulse from a laser-driven crystal}

Let us now consider Eq.~(\ref{Prob_General_coherent}) describing the interaction of an electronic system with a coherent x-ray pulse. Evaluating the integrals $\int d^3 re^{i \mathbf q_x\cdot\br}\widetilde M_{F_0I_0}(\mathbf r,\Delta\mu)$ for a crystal in Appendix \ref{App_Fourier}, we obtain that Eq.~(\ref{Prob_General_coherent}) can be represented as a sum of three terms 
\begin{widetext}
\begin{align}
P(\mathbf q_x) = &\sum_{\mathbf G}\theta(\mathbf q_x-\mathbf G) \Biggl[\sum_{\overline F_0,I_0,\Delta\mu,\Delta\mu'}|\mathcal C_{I_0}|^2\widetilde T_{\overline F_0I_0\Delta\mu'}^*(\mathbf G)\widetilde T_{\overline F_0I_0\Delta\mu}(\mathbf G)\nonumber\\
&\quad\qquad\qquad\qquad+\sum_{\overline F_0,I_0,K_0\neq I_0,\Delta\mu,\Delta\mu'}  \mathcal C^*_{K_0}\mathcal C_{I_0}\,\widetilde  T^*_{\overline F_0K_0\Delta\mu'}(\mathbf G)\widetilde T_{\overline F_0I_0\Delta\mu}(\mathbf G)\Biggr]
\label{Prob_General_coherent_crystal}\\
&+\sum_{F_0,I_0,K_0,\Delta\mu,\Delta\mu'}  \mathcal C^*_{K_0}\mathcal C_{I_0}\widetilde  T^*_{F_0K_0\Delta\mu'}(\mathbf q_x)\widetilde T_{F_0I_0\Delta\mu}(\mathbf q_x)\nonumber
\end{align}
with
\begin{align}
\widetilde T_{F_0I_0\Delta\mu}(\mathbf q_x) = \widetilde{\mathcal E}_x (\Omega_{F_0I_0}+\Delta\mu\omega)\int d^3 r e^{i\mathbf q_x\cdot\mathbf r}\widetilde M_{F_0I_0}(\br,\Delta\mu),
\end{align}
\end{widetext}
where a many-body Floquet eigenstate $F_0$ is such that it is obtained from a state $I_0$ by replacing a function $|\phi_{i,\k}\ra$ for $|\phi_{f,\k-\mathbf q_x+\mathbf G}\ra$ for some $\k$ and $\mathbf G$, or is equal to $I_0$ resulting in $\widetilde T_{I_0I_0\Delta\mu}(\mathbf q_x\neq \mathbf G)= 0$. In \eq{Prob_General_coherent_crystal}, the summation over states $\overline F_0$ runs through the Floquet eigenstates comprising the state of the light-dressed crystal $|\Psi_0,t\ra$ ($\mathcal C_{\overline F_0}\neq 0$) leading to quasielastic scattering, and the summation over states $F_0$ runs through the eigenstates that are not ($\mathcal C_{F_0}=0$) leading to inelastic scattering. We took into account that, in practice, the Dirac delta functions $\delta(\mathbf q_x-\mathbf G)$ resulting from the integrals $\int d^3 re^{i \mathbf q_x\cdot\br}\widetilde M_{\overline F_0I_0}(\mathbf r,\Delta\mu)$ ({\it cf.}~Appendix \ref{App_Fourier}) must be convoluted with a detector response function of finite resolution turning to some continuous functions $\theta(\mathbf q_x-\mathbf G)$. 

The strength of the scattering signal in the vicinity of reciprocal lattice vectors $\mathbf G$ is given by the two terms in the squared brackets in \eq{Prob_General_coherent_crystal} and the third term at $\mathbf q_x=\mathbf G$. The first term in \eq{Prob_General_coherent_crystal} is due quasielastic transitions from initial states $I_{\Delta\mu}$ to final states $F_{\Delta\mu'}$, leading to a signal centered at scattering energies $\omega_{\k_\s}=\omega_{x\i}+(\Delta\mu-\Delta\mu')\omega+E_{I_0}-E_{F_0}$. A contribution due to a transition from an initial state $I_{\Delta\mu}$ to a final state $I_{\Delta\mu'}$ leads to a signal centered at scattering energies $\omega_{\k_\s}=\omega_{x\i}+(\Delta\mu-\Delta\mu')\omega$. Please note that it cannot be distinguished from the contribution due to a transition from a different initial state $K_{\Delta\mu}$ to a final state $K_{\Delta\mu'}$. The second term in \eq{Prob_General_coherent_crystal} is due to the interference terms between quasielastic transitions from initial states that belong to a different family of Floquet replica states. The third contribution to the scattering signal at reciprocal lattice vectors $\mathbf G$ is due to inelastic transitions from initial states $I$ and $K$ to final states $F$, which differ from them by a single function occupied at some point $\k+\mathbf G$, but not at $\k$. It is nonzero even for x-ray scattering from driven electronic systems in an initial state described by a single series with $\mathcal C_{K_0}=\delta_{I_0,K_0}$ in \eq{Eq_wavepacket}. Analogously to x-ray scattering from stationary systems, the contributions due to quasielastic scattering in the case of $\Delta\mu=\Delta\mu'=0$ and $\overline F_0=I_0$ in the first term would dominate over inelastic contributions given by the third term. However, if $\Delta\mu$ and $\Delta\mu'$ are nonzero, the contributions due to quasielastic scattering could be comparable to or even smaller than the inelastic contributions described by the third term in \eq{Prob_General_coherent_crystal}. The reason for this is that $\widetilde T_{I_0I_0\Delta\mu}$ is determined by the sum $\sum_{i,\k} \sum_{m,m',\mu}c^{i*}_{m',\k,\mu+\Delta\mu}c^i_{m,\k,\mu} \int\limits_{V_{\text{cell}}} d^3r u_{m'\k}^\dagger(\br) u_{m\k}(\br)e^{i\mathbf G\cdot\mathbf r}$, which could be smaller than a single term $c^{f*}_{m',\k,\mu+\Delta\mu}c^i_{m,\k,\mu} \int\limits_{V_{\text{cell}}} d^3r u_{m'\k}^\dagger(\br) u_{m\k}(\br)e^{i\mathbf G\cdot\mathbf r}$ for $\Delta\mu\neq 0$, if the integrals $\int\limits_{V_{\text{cell}}} d^3r u_{m'\k}^\dagger(\br) u_{m\k}(\br)e^{i\mathbf G\cdot\mathbf r}$ do not vary much for $m$ and $m'$, due to the orthonormality of $c^i_{m,\k,\mu}$ coefficients: $\sum_{m,\mu,\k}c^{i*}_{m,\k,\mu+\Delta\mu}c^i_{m,\k,\mu} = \sum_{m,\mu,\k}c^{i_{\Delta\mu}*}_{m,\k,\mu}c^{i_0}_{m,\k,\mu}=\delta_{\Delta\mu,0}$.  Contributions from lattice disorder would additionally smear out the scattering signal at reciprocal lattice vectors $\mathbf G$.

It follows from \eq{Prob_General_coherent_crystal} that a scattering signal at reciprocal lattice vectors $\mathbf G$ does not automatically provide quasielastic scattering. If the bandwidth of the x-ray pulse is more narrow than energy differences between any states $I_0$, $K_{\Delta\mu}$ and $F_{\Delta\mu'}$ for any $\Delta\mu$ and $\Delta\mu'$, then the interference terms in the second term of \eq{Prob_General_coherent_crystal} would be zero. But the inelastic contributions to the scattering signal at the $\mathbf G$ vectors given by the third term would still remain. They can be separated from the quasielastic contributions in the first term only by the spectroscopy of the scattered photons.

\subsection{Quasielastic scattering from a laser-driven crystal in a state described by physically equivalent Floquet states}
\label{Subsec_quasieel_crystal}

Let us consider a laser-driven crystal prepared in a state $|\Psi_0,t\ra$ described by a single series $|\Theta_{I_0},t\ra$ of physically equivalent Floquet states, which corresponds to the electronic wave function of a crystal given by a superposition $|\Psi_{\el},t \ra= \sum_{\mu,n}C^{I_0}_{n,\mu}e^{-i\mu\omega}|\Phi_n\ra$ evolving in time periodically with the frequency $\omega$. In this case, the probability of quasielastic scattering by a perfectly coherent nonresonant high-energy x-ray probe pulse can be represented as 
\begin{align}
&P_{qe}(\mathbf q_x) = \sum_{\Delta\mu,\mathbf G}\widetilde P_{qe}(\mathbf G,\Delta\mu)\delta(\mathbf q_x-\mathbf G)\nonumber\\
&\qquad\qquad\qquad\times|\widetilde{\mathcal E}_x (\omega_\ks-\omega_{x\i}+\Delta\mu\omega)|^2\label{Eq_Bragg_peaks},\\
&\widetilde P_{qe}(\mathbf G,\Delta\mu) \propto\lf|\int d^3 r e^{i\mathbf G\cdot\mathbf r}\widetilde\rho(\br,\Delta\mu)\rt|^2.\nonumber
\end{align}
According to this expression, a scattering signal in this case is a series of Bragg peaks at crystal reciprocal lattice vectors $\mathbf G$ at scattering energies $\omega_{x\i}-\Delta\mu\omega$, which we will refer to as $\Delta\mu$-th order Bragg peaks. The intensity of the $\Delta\mu$-th order Bragg peak, $\mathcal I_{\Delta\mu}(\mathbf G)$, is proportional to $\widetilde P_{qe}(\mathbf G,\Delta\mu)$ and is given by the $\mathbf G$-th amplitude of the spatial Fourier transform of the $\Delta\mu$-th amplitude of the time-dependent density of a crystal $\rho(\br,t) = \sum_{\Delta\mu}e^{i\Delta\mu\omega t}\widetilde\rho(\br,\Delta\mu)$. 

This conclusion goes in line with the experiment by Glover {\it et al.~}in Ref.~\onlinecite{GloverNature12}, who have observed x-ray and optical wave mixing in a diamond crystal. In their experiment, a diamond sample was simultaneously illuminated by an x-ray and an optical pulse. They have observed a signal at a scattering vector $\mathbf q_x = \mathbf G+\boldsymbol\kappa_e$, where $\boldsymbol\kappa_e$ is the wave vector of the optical pulse, accompanying the diamond Bragg peak at $\mathbf G=(1,1,1)$. In that experiment, the incident x-ray energy at 8 keV had a bandwidth of approximately 1 eV, set by a Si$(1,1,1)$ double monochromator, and which is less than the 1.55 eV laser photon energy. X-ray photons at the sum frequency were detected after a Si $(2,2,0)$ channel cut analyzer and were distinguished from the elastic scattering given by $P(\omega_{\k_\s}=\omega_{x\i})$ in energy and both the phase matching and emission angle. The results were interpreted as arising due to the inelastic scattering of the incident x-rays from the $(1,1,1)$ Fourier component of the optically induced currents, from which they extract a corresponding change in valence charge density.

We have a slightly different interpretation for the observed signal in the experiment of Ref.~\onlinecite{GloverNature12} and suggest that it is the first-order Bragg peak given by the probability $\widetilde P_{qe}(\mathbf G,1)$ in Eq.~(\ref{Eq_Bragg_peaks}). The probability $\widetilde P_{qe}(\mathbf G,1)$ is determined by the amplitudes $\widetilde \rho(\br,1)$, which are much larger than other higher-order amplitudes, $\widetilde \rho(\br,\Delta\mu)$ for $|\Delta\mu|>1$ at the condition of their experiment, where the pump pulse interacted with the crystal in a perturbative regime. Since, in this regime, $\widetilde \rho(\br,0)$ gives approximately the unperturbed density of the crystal, $\widetilde \rho(\br,1)$ does describe the optically-induced change of the electron density. However, electrons brought to conduction bands by the pump pulse also contribute to $\widetilde \rho(\br,1)$ [{\it cf.}~\eq{rhoI0I0viaOneBody}] and, thus, the signal is connected to the total change of the electron density.

Eq.~(\ref{Eq_Bragg_peaks}) does not describe the shift of a scattering vector by $\boldsymbol \kappa_0$ relative to a reciprocal lattice vector $\mathbf G$. This discrepancy is due to the dipole approximation to the interaction between the crystal and the driving electromagnetic field. This assumption results in the approximated time-dependent electron density with the same spatial periodicity as the stationary electron density of a crystal. Please notice that this discrepancy follows only from the approximation to the electron density, but not from the theory describing the interaction with a nonresonant x-ray probe pulse.



\section{Non-resonant x-ray scattering from a light-driven $\mathrm{\mathbf{MgO}}$ crystal}
\label{Section_MgO}

We illustrate our study with a calculation of the probability of nonresonant x-ray scattering by a temporally periodic coherent x-ray pulse from the cubic wide-bandgap crystal MgO driven by an intense infrared laser pulse of the photon energy $\omega = 1.55$ eV in a nonlinear regime. Recently, it was demonstrated that HHG is strongly sensitive to the atomic-scale structure of MgO, which was proposed as a possible probe of electron dynamics driven by an electromagnetic pulse in a crystal \cite{YouNature16}. Therefore, we have chosen a similar regime for the interaction with the driving laser pulse for our calculation in order to determine what new insights on electron dynamics of a light-driven crystal nonresonant x-ray scattering can provide.

\begin{figure}[t]
\centering
\includegraphics[width=0.5\textwidth]{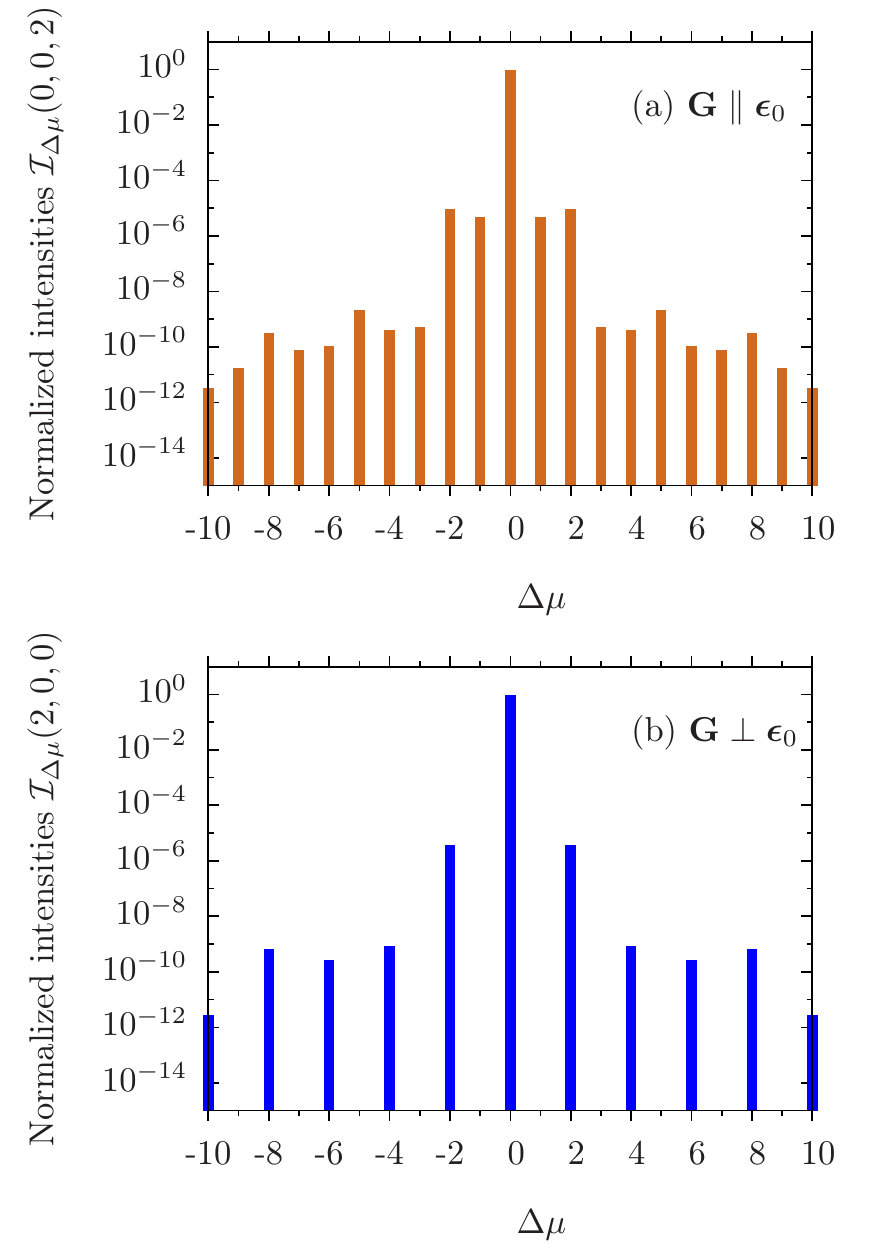}
\caption{Normalized intensities of Bragg peaks at scattering energies $\omega_{x\i}+\Delta\mu\omega$, $\mathcal I_{\Delta\mu}(\mathbf G)/\mathcal I_{0}(\mathbf G)$, depending on $\Delta\mu$ for (a) $\mathbf G  = (0,0,2)$ and (b) $\mathbf G = (2,0,0)$.}
\label{Fig_BraggsatCIntensity}
\end{figure}

We calculated the Bloch functions within density functional theory with the ABINIT software package \cite{Gonze16,*Gonze09,*Gonze05} using Troullier-Martins pseudopotentials \cite{Troullier-Martins_Pseudpotentials}. The calculated Bloch functions were used as basis functions [{\it cf.}~Eq.~(\ref{Eq_basis_set_onebody})], which were then used to diagonalize the Hamiltonian $\Hdel$ [Eqs.~(\ref{Eq_spatial_periodic_Ham1})-(\ref{Eq_spatial_periodic_Ham3})] at each $\k$ point. The resulting Floquet-Bloch eigenstates were substituted in Eq.~(\ref{Eq_Bragg_peaks}) to calculate the diffraction signal. According to our convergence study, the calculation of the diffraction signal is converged when a $24\times24\times24$ Monkhorst-Pack grid, sixteen conduction bands and $2\mu_{\text{max}}+1=81$ blocks of the Floquet Hamiltonian in \eq{Eq_FloquetMatrix} are taken into account for MgO crystal driven by a pump pulse of 1.55 eV photon energy and $2\times 10^{12}$ W/cm$^2$ intensity, which is the maximum intensity in our calculation. We also use these parameters for calculations at lower intensities, since the number of conduction bands and $\mu_{\text{max}}$ necessary for the convergence drops with decreasing intensity of the pump pulse. 


For simplicity, we took into account only four valence bands of MgO, but ignored the impact from inner-shells of the crystal, which does not influence calculation of the time-dependent density, but provides an additive to a diffraction signal. This results in our calculation being not quite precise, but still accurate enough to illustrate some features of nonresonant x-ray scattering from a driven crystal and demonstrate the feasibility of such a calculation.

\begin{figure}[t]
\centering
\includegraphics{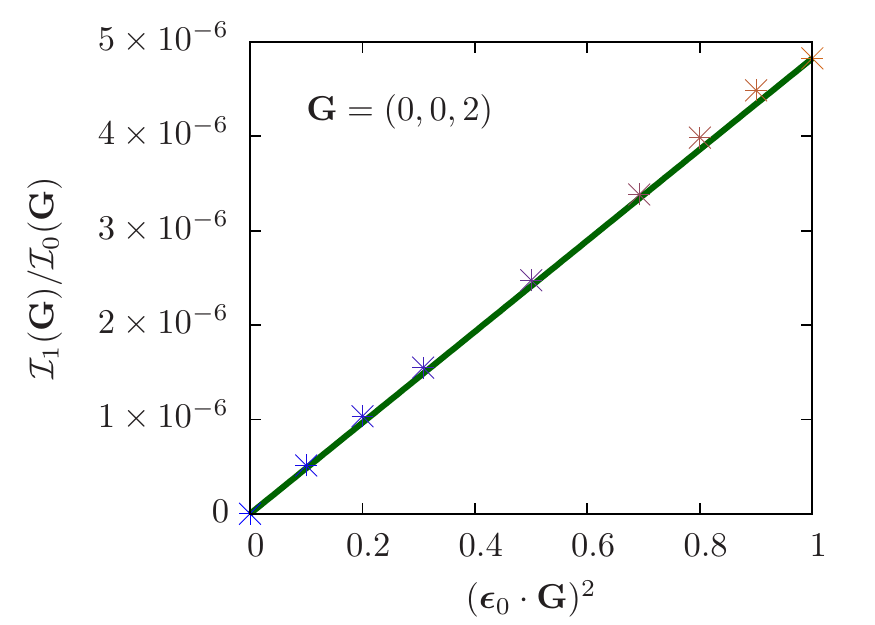}
\caption{Dependence of the normalized intensity of the first-order Bragg peak at $\mathbf G=(0,0,2)$, $\mathcal I_1(\mathbf G)/\mathcal I_0(\mathbf G)$, on polarization of the pump pulse $\boldsymbol\epsilon_0$.}
\label{Fig_polarization}
\end{figure}

We assume that the state of the laser-driven MgO crystal before the interaction with the probe pulse can be well described by a single series involving states $|\Psi_{I_{\Delta\mu}}\ra$ with the largest absolute values of projections on the ground electronic state of MgO. We also assume that a perfectly coherent nonresonant x-ray pulse of photon energy $\omega_{x\i}$ is used as a probe pulse. As described in the previous Section, the diffraction signal in this case consists of $\Delta\mu$-th order Bragg peaks at scattering energies $\omega_{x\i}-\Delta\mu\omega$ and at reciprocal lattice vectors $\mathbf G$, and their intensity is proportional to the squared spatial Fourier transform of the corresponding $\Delta\mu$-th amplitudes of the time-dependent electron density, $\lf|\int d^3 r e^{i\mathbf G\cdot\mathbf r}\widetilde\rho(\br,\Delta\mu)\rt|^2$.

Figure~\ref{Fig_BraggsatCIntensity} shows the intensities of the $\Delta\mu$-th order Bragg peaks normalized to the intensity of the zero-order Bragg peak at the scattering energy $\omega_{x\i}$, $\mathcal I_{\Delta\mu}(\mathbf G)/\mathcal I_{0}(\mathbf G)=\lf|\int d^3 r e^{i\mathbf G\cdot\mathbf r}\widetilde\rho(\br,\Delta\mu)\rt|^2/\lf|\int d^3 r e^{i\mathbf G\cdot\mathbf r}\widetilde\rho(\br,0)\rt|^2$. Here, we assume the pump pulse of $2\times 10^{12}$ W/cm$^2$ intensity polarized along the (0,0,1) direction. We consider two cases: vector $\mathbf G=(0,0,2)$ parallel to the pump-pulse polarization $\boldsymbol\epsilon_0$ [Fig.~\ref{Fig_BraggsatCIntensity}(a)] and vector $\mathbf G=(2,0,0)$ perpendicular to $\boldsymbol\epsilon_0$ [Fig.~\ref{Fig_BraggsatCIntensity}(b)]. As follows from the Figures, the ratios $\mathcal I_{\Delta\mu}(\mathbf G)/\mathcal I_{0}(\mathbf G)$ are lower than $10^{-5}$. Still, it should be experimentally feasible to observe the $\Delta\mu$-th order Bragg peaks, for example, following the technique of Glover {\it et al.~}in Ref.~\onlinecite{GloverNature12}, who have observed the first-order Bragg peak.

The Bragg peaks at the vector $\mathbf G$ parallel to the pump-pulse polarization are of both even and odd orders, whereas the Bragg peaks at the vector $\mathbf G$ perpendicular to the pump-pulse polarization $\boldsymbol\epsilon_0$ appear only at even orders. This demonstrates a strong anisotropy of odd-order amplitudes $\widetilde\rho(\br,\Delta\mu)$ of the time-dependent electron density in MgO with respect to the polarization of the laser pulse driving electron dynamics. The observation that $\int d^3 r e^{i\mathbf G\cdot\mathbf r}\widetilde\rho(\br,\Delta\mu)$ at the vector $\mathbf G\perp\boldsymbol\epsilon_0$ is zero for odd $\Delta\mu$ may be related to the effect that HHG spectra of MgO contain only odd-order harmonics \cite{YouNature16}, which also become zero in the scattering signal at $\mathbf G\perp\boldsymbol\epsilon_0$ in our calculations. The latter phenomenon was attributed to a highly directional field-induced nonlinear current \cite{YouNature16}. Since only odd harmonics are observed in the HHG spectra of MgO, the current induced by the pump pulse may not influence even-order laser-driven electronic properties, and, thus, the even-order amplitudes of the electron density remain unaffected by the direction of the field-induced current. A more detailed investigation of this phenomenon is required for its precise interpretation, which is beyond the scope of this paper.

\begin{figure}[t]
\centering
\includegraphics[width=0.5\textwidth]{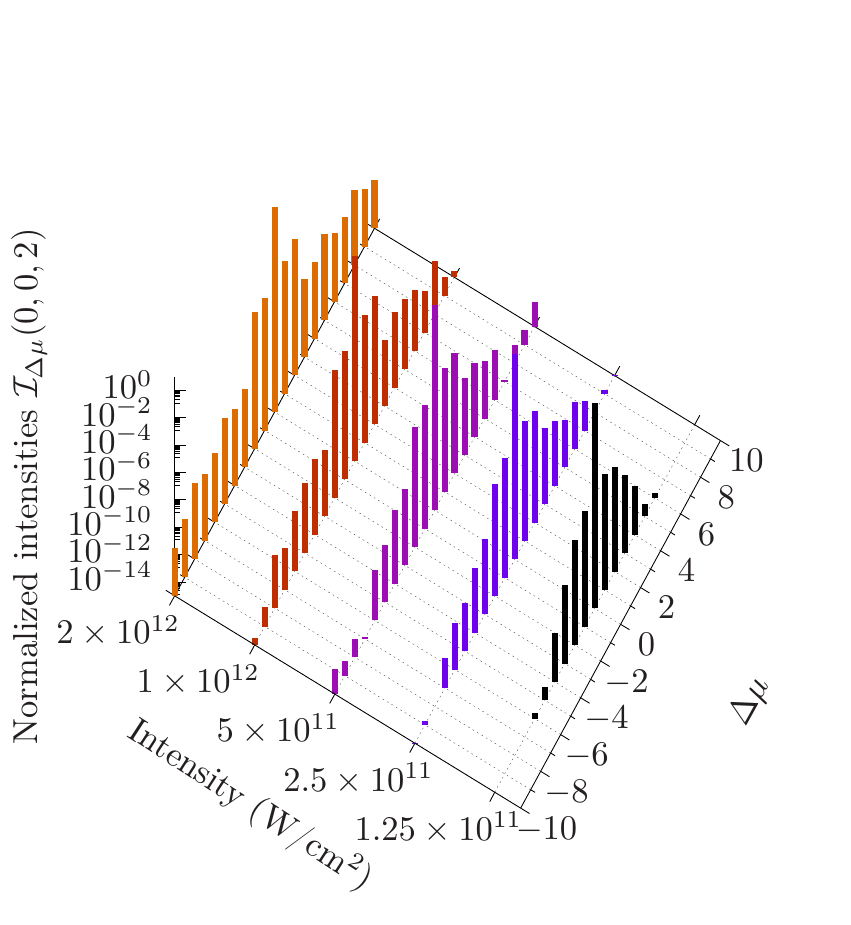}
\caption{Normalized intensities of Bragg peaks $\mathcal I_{\Delta\mu}(0,0,2)/\mathcal I_{0}(0,0,2)$ depending on $\Delta\mu$ at different intensities of the pump pulse.}
\label{Fig_intensity}
\end{figure}

Figure \ref{Fig_polarization} shows the dependence of the first-oder Bragg peak intensity at $\mathbf G=(0,0,2)$ on the polarization of the pump pulse. It follows from the figure that the intensity of the first-order Bragg peak is proportional to $(\boldsymbol{\epsilon}_{0}\cdot \mathbf G)^2$. Thus the spatial Fourier transform of the amplitude $\widetilde\rho(\br,1)$ of the time-dependent density of MgO has a linear dependence on $(\boldsymbol{\epsilon}_{0}\cdot \mathbf G)$. This result agrees with the experiment by Glover {\it et al.~}in Ref.~\onlinecite{GloverNature12} mentioned above, who have observed the same dependence of the signal intensity on the polarization of an optical pulse.

The experiment by You {\it et al.}~in Ref.~\onlinecite{YouNature16} has also demonstrated the anisotropy of the interaction between the driving pulse and the MgO crystal. However, their experiment did not reveal selection rules and polarization dependence of the time-dependent electron density. In addition, only odd harmonics in the HHG spectra of MgO appear, which makes any even-order effects undetectable by this technique. Thus, nonresonant x-ray scattering providing direct information about the time-dependent electron density can be used as a complementary technique to probe electron dynamics in crystals during their interaction with an intense laser pulse.


\begin{figure}[t]
\centering
\includegraphics[width=0.5\textwidth]{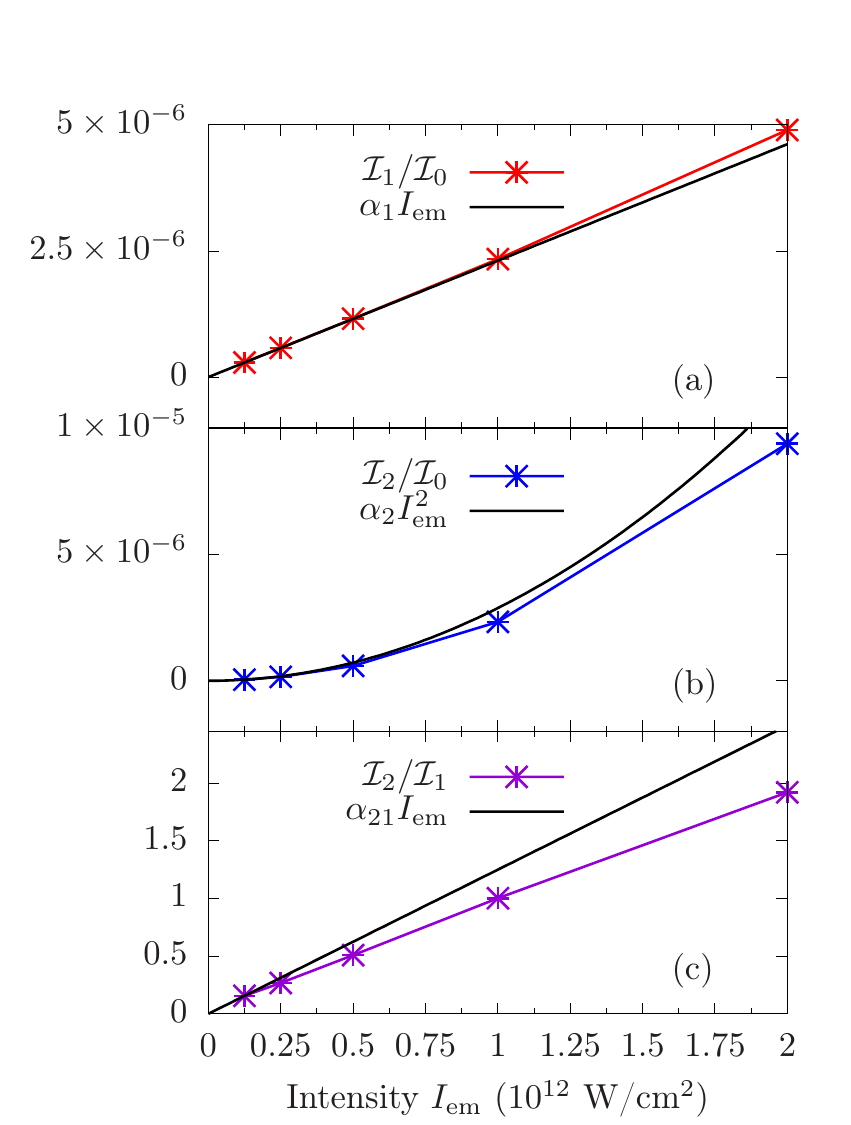}
\caption{Comparison of ratios of Bragg-peak intensities as functions of the pump-pulse intensity $I_{\text{em}}$ at $\mathbf G = (0,0,2)$ with functions of $I_{\text{em}}$: (a) $\mathcal I_1(\mathbf G)/\mathcal I_0(\mathbf G)$ with $\alpha_1 I_{\text{em}}$; (b) $\mathcal I_2(\mathbf G)/\mathcal I_0(\mathbf G)$ with $\alpha_2 I_{\text{em}}^2$; (c) $\mathcal I_2(\mathbf G)/\mathcal I_1(\mathbf G)$ with $\alpha_{21} I_{\text{em}}$. $\alpha_1$, $\alpha_2$ and $\alpha_{21}$ are chosen such that the ratios of the Bragg-peak intensities and the corresponding functions of $I_{\text{em}}$ coincide at $I_{\text{em}}=1.25\times10^{11}$ W/cm$^2$.}
\label{Fig_pintensity}
\end{figure}

In Fig.~\ref{Fig_intensity}, we study how the distribution of intensities $\mathcal I_{\Delta\mu}(0,0,2)$ changes depending on the intensity of the driving pulse polarized along $(0,0,1)$. When the pump-pulse intensity is $1.25\times10^{11}$ W/cm$^2$ (the plot with black columns in Fig.~\ref{Fig_intensity}), the intensities of $\Delta\mu$-th order Bragg peaks monotonically decrease with increasing $|\Delta\mu|$, which can be described within low-order perturbation theory. In contrast, the plot with orange columns in Fig.~\ref{Fig_intensity} corresponding to the pump-pulse intensity of $I_{\text{em}}=2\times10^{12}$ W/cm$^2$ [the same plot as in Fig.~\ref{Fig_BraggsatCIntensity}(a)] shows nonuniform distribution of intensities $\mathcal I_{\Delta\mu}$. For example, the intensity of the second-order Bragg peak is higher than the intensity of the first-order Bragg peak on this plot meaning that $\lf|\int d^3 r e^{i\mathbf G\cdot\mathbf r}\widetilde\rho(\br,2)\rt|>\lf|\int d^3 r e^{i\mathbf G\cdot\mathbf r}\widetilde\rho(\br,2)\rt|$ at $I_{\text{em}}=2\times10^{12}$ W/cm$^2$. This indicates a nonperturbative nature of the interaction between the pump pulse of $2\times10^{12}$ W/cm$^2$ intensity and the MgO crystal. Studying the other distributions, one can observe how the time-dependent electron density changes with increasing pump-pulse intensity making a transition from a low-order perturbative to a nonperturbative regime. For example, Fig.~\ref{Fig_pintensity} shows the ratios $\mathcal I_1(\mathbf G)/\mathcal I_0(\mathbf G)$, $\mathcal I_2(\mathbf G)/\mathcal I_0(\mathbf G)$ and $\mathcal I_2(\mathbf G)/\mathcal I_1(\mathbf G)$ at $\mathbf G=(0,0,2)$ as functions of the pump-pulse intensity. In a perturbative regime of interaction between the pump pulse and the MgO crystal, $\lf|\int d^3 r e^{i\mathbf G\cdot\mathbf r}\widetilde\rho(\br,2)\rt|^2\propto\mathcal I_2(\mathbf G)/\mathcal I_0(\mathbf G)$ should be a quadratic function of $I_{\text{em}}$, and $\lf|\int d^3 r e^{i\mathbf G\cdot\mathbf r}\widetilde\rho(\br,1)\rt|^2\propto\mathcal I_1(\mathbf G)/\mathcal I_0(\mathbf G)$ and $\mathcal I_2(\mathbf G)/\mathcal I_1(\mathbf G)$ should be linear functions of $I_{\text{em}}$. This is true only below $I_{\text{em}} = 5\times 10^{11}$ W/cm$^2$ indicating a transition to a nonperturbative regime of interaction at approximately this pump-pulse intensity.

The linear dependence of the relative intensity of the first-order Bragg peak on the pump-pulse intensity, when the pump-pulse intensity is below $5\times 10^{11}$ W/cm$^2$, also agrees with the experiment of Glover {\it et al.~}in Ref.~\onlinecite{GloverNature12}. In this experiment, they used a pump-pulse of the intensity $I_{\text{em}}\approx1.5\times10^{10}$ W/cm$^2$. The relative intensity of the peak $\mathcal I_1(\mathbf G)/\mathcal I_0(\mathbf G)$, which the authors referred to as the SFG efficiency, varied linearly with the intensity. The order of the effect also agrees with the experiment. In our case, the relative intensity of the first-order Bragg peak is $2.9\times10^{-7}$ at $I_{\text{em}}=1.25\times10^{11}$ W/cm$^2$ for MgO. In the experiment, it is $\approx3\times10^{-7}$ at $I_{\text{em}}\approx1.5\times10^{10}$ W/cm$^2$ for diamond.

  

\section{Conclusions}

In this manuscript, we described nonresonant x-ray scattering from an electronic system in the presence of a single-mode electromagnetic pulse with the frequency $\omega$ driving electron dynamics in this system. The driving field brings the electronic system to a state $|\Psi_0,t\ra$, which is a superposition of Fourier series involving physically equivalent Floquet states $K_{\Delta\mu}$ with eigenenergies shifted by multiples of $\omega$. If the driven electronic state is prepared in such a way that its state is described by a single series, then its electronic properties change periodically with the frequency $\omega$. We took into account that a nonresonant x-ray probe pulse can induce transitions from the states $K_{\Delta\mu}$ comprising $|\Psi_0,t\ra$ to any possible final Floquet states $F_{\Delta\mu'}$. We derived a general expression for the scattering probability of a nonresonant hard-x-ray probe pulse of arbitrary coherence properties and a duration valid for both time-unresolved and -resolved measurements, and considered particular cases following from this expression.

We obtained that the probability of x-ray scattering in general is not connected to the time-dependent electron density of the driven electronic system. In particular, it cannot be connected to the time-dependent density of a driven electronic system in a state involving physically inequivalent Floquet states, when its electronic properties do not evolve periodically with the frequency $\omega$, under any conditions. 



If the state of the driven electronic system does evolve periodically with the frequency $\omega$, then it is possible to connect the scattering signal to the time-dependent electron density in the following case. If the bandwidth of the probe x-ray pulse is smaller than any energy differences between states $K_{\Delta\mu}$ comprising the state of the driven electronic system $|\Psi_0,t\ra$ and possible final Floquet states $F_{\Delta\mu'}$ for any $\Delta\mu$ and $\Delta\mu'$, then it is possible to separate from the total signal the contribution due to quasielastic transitions, {\it i.e.~}transitions with final states being the states comprising $|\Psi_0,t\ra$, by the spectroscopy of scattered photons. In this case, the probability of quasielastic scattering is connected to the time-dependent electron density. In particular, if the probe x-ray pulse is spatially uniform and coherent, the scattering signal would include a series of peaks at scattering energies $\omega_{x\i}-\Delta\mu\omega$ with amplitudes determined by the spatial Fourier transform of the corresponding amplitude $\widetilde\rho(\br,\Delta\mu) $ of the time-dependent electron density given by $\rho(\br,t) = \sum_{\Delta\mu}e^{i\Delta\mu\omega t}\widetilde\rho(\br,\Delta\mu)$.

In contrast to measurements of stationary electronic systems, when inelastic contributions are negligible in comparison to elastic ones, inelastic contributions can be comparable to or larger than quasielastic contributions related to $\widetilde\rho(\br,\Delta\mu\neq0)$ and cannot be neglected in the case of a driven electronic system. In a general case, and, particularly, in the case of an ultrashort probe x-ray pulse, it may not be possible to spectroscopically distinguish between inelastic and quasielastic contributions to the scattering probability. If the bandwidth of a probe pulse is not narrow enough, quasielastic contributions intermix with inelastic ones in the spectrum and cannot be factored out.  Applying ultrashort x-ray pulses in experiments aimed to probe the time-dependent electron density of a driven electronic system, one has to take this aspect into account.

We showed how our study can be applied to the calculation of nonresonant x-ray scattering from a crystal driven by an electromagnetic pulse. In particular, we showed that quasielastic scattering of a spatially uniform coherent narrow-bandwidth probe x-ray pulse from driven crystals with electronic properties periodically evolving with the frequency $\omega$ results in the appearance of $\Delta\mu$-th order Bragg peaks with intensities proportional to $\lf|\int d^3 r e^{i\mathbf G\cdot\mathbf r}\widetilde\rho(\br,\Delta\mu)\rt|^2$ at scattering energies $\omega_{x\i}-\Delta\mu\omega$. For these conditions, we illustrated some features of nonresonant x-ray scattering from a laser-driven crystal and the information it can provide by considering a MgO crystal driven by a laser pulse in a strongly nonlinear regime. Nonresonant x-ray scattering revealed special selection rules of the interaction between the driving laser pulse and MgO crystal resulting in the strong anisotropy of odd-order amplitudes of the time-dependent electron density.

To sum up, we showed how a nonresonant x-ray probe pulse interacts with a laser-driven electronic system and discussed particular cases, when its time-dependent electron density can be imaged by means of x-ray scattering. The ability to follow electronic dynamics in laser-driven electronic systems opens up opportunities for better understanding and control of the way how electronic properties of such systems are modified by a driving electromagnetic pulse.

\section*{Acknowledgment}
We would like to acknowledge the Stephenson Distinguished Visitor  Programme  (DESY  Photon  Science),  Year  2017,  for
supporting the project. D. A. Reis acknowledges the support by the AMOS program within the Chemical Sciences Division of the Office of Basic Energy Sciences, Office of Science, U.S. Department of Energy.


\widetext
\appendix
\section{Time-dependent electron density}
\label{App_density0}

The time-dependent electron density $\rho(t)$ is given by
\begin{align}
\rho(\br,t)&=\la \Psi_{0},t|\hat\psi^\dagger(\br)\hat\psi(\br)|\Psi_0,t\ra\label{density_app}\\
& = \sum_{I_0,K_0}\mathcal C_{K_0}^*\mathcal C_{I_0}e^{-i(E_{I_0}-E_{K_0})t}\sum_{\mu,\mu',n,n'}e^{-i(\mu-\mu')\omega t}C^{K_0^*}_{n',\mu'}C^{I_0}_{n,\mu}\la\Phi_{n'}|\hat\psi^\dagger(\br)\hat\psi(\br)|\Phi_{n}\ra
.\nonumber
\end{align}
Representing the sum over $\mu$ and $\mu'$ as
\begin{align}
\sum_{\mu,\mu'}e^{-i(\mu-\mu')\omega t}C^{K_0^*}_{n',\mu'}C^{I_0}_{n,\mu}=\sum_{\mu,\Delta\mu}C^{K_0^*}_{n',\mu+\Delta\mu}C^{I_0}_{n,\mu}e^{i\Delta\mu\omega t},
\end{align}
we obtain
\begin{align}
\rho(\br,t)&= \sum_{I_0,K_0}\mathcal C_{K_0}^*\mathcal C_{I_0}e^{-i(E_{I_0}-E_{K_0})t}\sum_{\Delta\mu}e^{-i\Delta\mu\omega t}\widetilde\rho_{K_0I_0}(\br,\Delta\mu),\nonumber
\end{align}
where 
\begin{align}
\rho_{K_0I_0}(\br,\Delta\mu) = C^{K_0^*}_{n',\mu+\Delta\mu}C^{I_0}_{n,\mu}\la\Phi_{n'}|\hat\psi^\dagger(\br)\hat\psi(\br)|\Phi_{n}\ra.
\end{align}

\section{Scattering probability}
\label{App_scatt_prob}

The interaction Hamiltonian entering Eq.~(\ref{Psi1_Eq}) in the main text can be represented as
\begin{align}
\hat H_{\text{int}x} = \frac{1}{c^2}\sum_{\ka_1,s_1}\sum_{\ks,s_\s}\sqrt{\frac{2\pi c^2}{V\omega_{x\ka_1}}} \sqrt{\frac{2\pi c^2}{V\omega_{x\ka_s}}}\hat a_{\ks,s_\s}^\dagger\hat a_{\ka_1,s_1}(\boldsymbol\epsilon_{x\ka_1,s_1}\cdot \boldsymbol\epsilon^*_{x\ks,s_\s})e^{i(\ka_1-\ks)\cdot\mathbf r}.
\end{align}
Substituting the interaction Hamiltonian into Eq.~(\ref{rho1_Eq}), we obtain that the following function enters the expression for the scattering probability
\begin{align}
&\sum_{\ka_1,\ka_2, s_1,s_2}\frac{2\pi\sqrt{\omega_{x\ka_1}\omega_{x\ka_2}}}{V}\sum_{\{n_x\},\{\widetilde n_x\}}\rho ^x_{\{n_x\},\{\widetilde n_x\}} \label{Eq_App_G_long}\\
&\times \sum_{\{n'_x\}}\la\{n'_x\}| \hat a_{\ks,s_\s}^\dagger\hat a_{\ka_1,s_1} |\{n_x\}\ra\la\{\widetilde n_x\}| \hat a^\dagger_{\ka_2,s_2}\hat a_{\ks,s_\s} |\{n'_x\}\ra e^{-i\omega_{x\ka_1}t_1}e^{i\omega_{x\ka_2}t_2}e^{i\ka_1\cdot\mathbf r_1}e^{-i\ka_2\cdot\mathbf r_2}
\nonumber.
\end{align}
This function is the first-order radiation correlation function $ G^{(1)}(\br_2,t_2,\br_1,t_1)$, since applying that $\sum_{\{n'_x\}}\hat a_{\ks,s_\s} |\{n'_x\}\ra\la\{n'_x\}| \hat a_{\ks,s_\s}^\dagger=1$, \eq{Eq_App_G_long} reduces to
\begin{align}
\sum_{\ka_1,\ka_2, s_1,s_2}\frac{2\pi\sqrt{\omega_{x\ka_1}\omega_{x\ka_2}}}{V}\operatorname{Tr}\bigl[\hat\rho ^x\hat a_{\ka_1,s_1}  \hat a^\dagger_{\ka_2,s_2}e^{-i\omega_{x\ka_1}t_1}e^{i\omega_{x\ka_2}t_2}e^{i\ka_1\cdot\mathbf r_1}e^{-i\ka_2\cdot\mathbf r_2}\bigr] = 
G^{(1)}(\br_2,t_2,\br_1,t_1),
\end{align}
where $\hat\rho ^x$ is the density matrix of the x-ray field. Thus, we obtain from \eq{Prob_denmatrix} that
\begin{align}
P(\omega_{\ks})
=&\frac{2\pi}{V\omega_\ks\omega_{x\i}^2}\sum_{s_\s,F}\bigl|(\boldsymbol\epsilon_{x\i}\cdot \boldsymbol\epsilon^*_{x\ks,s_\s})\bigr|^2 \int_{-\infty}^{+\infty} dt_1\int_{-\infty}^{+\infty} dt_2\\
&\qquad\times\int d^3 r_1\int d^3 r_2 G^{(1)}(\br_2,t_2,\br_1,t_1) \overline M_{F}(\br_1,t_1) \overline M^*_{F}(\br_2,t_2)e^{i\omega_\ks (t_1-t_2)-i\ks\cdot(\br_1-\br_2)},\nonumber
\end{align}
where
\begin{align}
\overline M_{F}(\br,t) = e^{iE_{F}t}
\la\Psi_{F}|\hat\psi^\dagger(\mathbf r)\hat\psi(\mathbf r) | \Psi_0,t\ra.\label{Eq_App_TF0}
\end{align}

Each final Floquet state $F$ is a member of a family of replica states according to Eq.~(\ref{EqReplicaStates}), $\Psi_{F_{\Delta\mu'}} =\sum_{n',\mu}C_{n',\mu+\Delta\mu'}^{F_0}|\Phi_{n'}\ra|N-\mu\ra$, with energies $E_{F_{\Delta\mu'}} = E_{F_0}+\Delta\mu'\omega$. Thus, we replace the sum over $F$ in the expression $\sum_F \overline M_{F}(\br_1,t_1)\overline M^*_{F}(\br_2,t_2)$ with $\sum_{\Delta\mu',F_{0}} M_{F_{\Delta\mu'}}(\br_1,t_1) M^*_{F_{\Delta\mu'}}(\br_2,t_2)$. Let us evaluate this sum 
\begin{align}
&\sum_{\Delta\mu',F_0}\overline M_{F_{\Delta\mu'}}(\br_1,t_1) \overline M_{F_{\Delta\mu'}}^*(\br_2,t_2) = \\
&= \sum_{\Delta\mu',F_0}e^{i(E_{F_0}+\Delta\mu'\omega)(t_1-t_2)}
\la\Psi_{0},t_2|\hat\psi^\dagger(\mathbf r_2)\hat\psi(\mathbf r_2) | \Psi_{F_\Delta\mu'}\ra\la\Psi_{F_\Delta\mu'}|\hat\psi^\dagger(\mathbf r_1)\hat\psi(\mathbf r_1) | \Psi_0,t_1\ra\nonumber\\
& =\sum_{\Delta\mu',F_0} e^{i(E_{F_0}+\Delta\mu'\omega)(t_1-t_2)}
\sum_{n,n',\mu,\mu'}C^{F_0^*}_{n',\mu'+\Delta\mu'}C^{F_0}_{n,\mu+\Delta\mu'}A_{N-\mu'}^*A_{N-\mu}e^{i(\mu\omega t_1-\mu'\omega t_2)}\nonumber\\
&\quad\times\la\Psi_{\text{el}},t_2|\hat\psi^\dagger(\mathbf r_2)\hat\psi(\mathbf r_2) | \Phi_{n'}\ra\la\Phi_{n}|\hat\psi^\dagger(\mathbf r_1)\hat\psi(\mathbf r_1) | \Psi_{\text{el}},t_1\ra,\nonumber
\end{align}
where we applied that
\begin{align}
\la\alpha,t_2|N-\mu'\ra\la N-\mu|\alpha,t_1\ra = A_{N-\mu'}^*A_{N-\mu}e^{i(-[N-\mu\omega] t_1-[N-\mu'\omega] t_2)}.
\end{align}
We now use the substitutions $\Delta\mu''=\Delta\mu'+\mu$ and $\Delta\mu'''=\Delta\mu'+\mu'$, and obtain
\begin{align}
&\sum_{\Delta\mu',F_0}\overline M_{F_{\Delta\mu'}}(\br_1,t_1) \overline M_{F_{\Delta\mu'}}^*(\br_2,t_2) = \\
&=\sum_{\Delta\mu'',\Delta\mu''',F_0} e^{iE_{F_0}(t_1-t_2)}e^{i(\Delta\mu''\omega t_1-\Delta\mu'''\omega t_2)}
\sum_{n,n'}C^{F_0^*}_{n',\Delta\mu'''}C^{F_0}_{n,\Delta\mu''}\sum_{\Delta\mu'}A_{N-\Delta\mu'''+\Delta\mu'}^*A_{N-\Delta\mu''+\Delta\mu'}\nonumber\\
&\quad\times\la\Psi_{\text{el}},t_2|\hat\psi^\dagger(\mathbf r_2)\hat\psi(\mathbf r_2) | \Phi_{n'}\ra\la\Phi_{n}|\hat\psi^\dagger(\mathbf r_1)\hat\psi(\mathbf r_1) | \Psi_{\text{el}},t_1\ra.\nonumber
\end{align}
The assumption that $\sum_{\Delta\mu'}A_{N-\Delta\mu'''+\Delta\mu'}^*A_{N-\Delta\mu''+\Delta\mu'}\approx 1$ independently of $\Delta\mu'''$ and $\Delta\mu''$ leads to
\begin{align}
\sum_{\Delta\mu',F_0}\overline M_{F_{\Delta\mu'}}(\br_1,t_1) \overline M_{F_{\Delta\mu'}}^\dagger(\br_2,t_2) = 
\sum_{F_0}M_{F_0}(\br_1,t_1) M^*_{F_0}(\br_2,t_2),
\end{align}
where 
\begin{align}
M_{F_0}(\br,t)= \sum_{I_0}\mathcal C_{I_0}e^{i(E_{F_0}-E_{I_0})t}\sum_{\Delta\mu}e^{i\Delta\mu\omega t}\widetilde M_{F_0I_0}(\br,\Delta\mu)
\end{align}
with
\begin{align}
\widetilde M_{F_0I_0}(\br,\Delta\mu) = \la\Psi_{F_{\Delta\mu}}|\hat\psi^\dagger(\br)\hat\psi(\br)|\Psi_0\ra.\label{Eq_App_MF0I0}
\end{align}
Thereby, $\widetilde M_{K_0I_0}(\br,\Delta\mu)=\widetilde\rho_{K_0I_0}(\br,\Delta\mu)$.

\section{Representation of functions $\widetilde M_{F_0I_0}$ via one-body Floquet states}
\label{App_MF0I0}

Let us evaluate functions $\widetilde M_{F_0I_0}$ for an electron system of noninteracting electrons. In this case, the many-body Hamiltonian of the system light and matter $\hat H_{\text{el-em}}$ can be written as a sum of independent one-body Hamiltonians. 

Let us consider a many-body solution for a many-electron system of the time-dependent Schr\"odinger equation, which, according to Eqs.~(\ref{Eq_Psi0Factorized})-(\ref{Eq_Theta}), can be represented as
\begin{align}
|\Psi_{\text{el}},t\ra = \sum_{I_0}\mathcal C_{I_0}e^{-iE_{I_0}t}|\Theta_{\text{el}I_0},t\ra.
\end{align}
Since $\mathcal C_{K_0}$ is determined by the boundary conditions, a time-dependent many-body function $|\Theta_{\text{el}I_0},t\ra$ is also a possible many-body solution to the time-dependent Schr\"odinger equation under certain boundary conditions. Let us assume that at $t=0$, it is given by a Slater determinant $|\Theta_{\text{el}I_0},0\ra = |\phi^{\el0}_{1,1},\cdots, \phi^{\el0}_{i,\k},\cdots\ra$, where $|\phi^{\el0}_{i,\k}\ra = \sum_{m,\mu}c^{i}_{m,\k,\mu}|\varphi_{m\k}\ra$. In the case of noninteracting electrons, the time evolution of this many-body wave function $|\Theta_{\text{el}I_0},t\ra$ can be represented as a Slater determinant $|\phi^{\el}_{1,1},\cdots, \phi^{\el}_{i,\k},\cdots\ra$, where $\phi^\el_{i,\k}(\br,t) = \sum_{m,\mu}c^{i}_{m,\k,\mu}e^{-i\mu\omega t}\varphi_{m\k}(\br)$ is a one-body solution of the time-dependent Schr\"odinger equation for the boundary condition $\phi^\el_{i,\k}(\br,0) = \phi^{\el0}_{i,\k}$. Therefore, we can obtain the matrix elements
\begin{align}
N_{F_0I_0}(\br,t)=\la \Theta_{\text{el}F_0},t|\hat\psi^\dagger(\br)\hat\psi(\br)|\Theta_{\text{el}I_0},t\ra
\end{align}
using the relations
\begin{align}
N_{I_0I_0}(\br,t) & = 
\sum_{i,\k}|\phi^\el_{i,\k}(\br,t)|^2\label{Eq_App_NI0I0}\\
& =\sum_{\Delta\mu}e^{i\Delta\mu\omega t} \sum_{m,m',\mu}c^{i*}_{m',\k,\mu+\Delta\mu}c^{i}_{m,\k,\mu}\varphi_{m'\k}^\dagger(\br)\varphi_{m\k}(\br),\nonumber
\end{align}
where the sum is over such $i$ and $\k$ that $|\phi^\el_{i,\k}\ra$ enters $|\Theta_{\text{el}I_0},t\ra$, and
\begin{align}
N_{F_0\neq I_0}(\br,t)& = 
\phi^{\el\dagger}_{f,\k'}(\br,t)\phi^\el_{i,\k}(\br,t)\label{Eq_App_NF0I0}\\ 
&=\sum_{\Delta\mu}e^{i\Delta\mu\omega t} \sum_{m,m',\mu}c^{f*}_{m',\k',\mu+\Delta\mu}c^{i}_{m,\k,\mu}\varphi_{m'\k'}^\dagger(\br)\varphi_{m\k}(\br)\nonumber,
\end{align}
which is nonzero, if the Slater determinant $|\Theta_{\text{el}F_0},t\ra$ can be obtained from $|\Theta_{\text{el}I_0},t\ra$ by replacing a function $\phi^\el_{i,\k}(\br,t)$ by $\phi^\el_{f,\k'}(\br,t)$.

Let us now express the matrix elements $N_{F_0I_0}$ via the functions $\widetilde M_{F_0I_0}(\br,\Delta\mu)$ in \eq{Eq_App_MF0I0}. Since $|\Theta_{I_0},t\ra=|\Theta_{\text{el}I_0},t\ra|\alpha,t\ra$, we apply that $N_{F_0I_0}=\la \Theta_{F_0},t|\hat\psi^\dagger(\br)\hat\psi(\br)|\Theta_{I_0},t\ra $. Then, we express $|\Theta_{I_0},t\ra$ via the Floquet eigenstates $|\Psi_{I_{\Delta\mu}}\ra$ as follows
\begin{align}
|\Theta_{I_0},t\ra& = \sum_{\Delta\mu,K_0}|\Psi_{K_{\Delta\mu}}\ra\la\Psi_{K_{\Delta\mu}}|\Theta_{I_0},t\ra\\
&=\sum_{\Delta\mu}|\Psi_{I_{\Delta\mu}}\ra\la\Psi_{I_{\Delta\mu}}|\Theta_{I_0},t\ra\nonumber\\
&=\sum_{\Delta\mu}B^{I_0}_{\Delta\mu}(t)|\Psi_{I_{\Delta\mu}}\ra\nonumber,
\end{align}
where, according to \eq{Eq_Theta},
\begin{align}
B^{I_0}_{\Delta\mu} =\sum_n\la\Psi_{I_{\Delta\mu}}|\Phi_n\ra|\alpha,t\ra\sum_{\mu}C^{I_0}_{n,\mu}e^{-i\mu\omega t}.
\end{align}
Thus,
\begin{align}
N_{F_0I_0}=\sum_{\Delta\mu,\Delta\mu'}B^{F_0^*}_{\Delta\mu'+\Delta\mu}B^{I_0}_{\Delta\mu}\la\Psi_{F_{\Delta\mu'}}|\hat\psi^\dagger(\br)\hat\psi(\br)|\Psi_{I_{0}}\ra.
\end{align}
We now consider the sum 
\begin{align}
\sum_{\Delta\mu}B^{F_0^*}_{\Delta\mu'+\Delta\mu}B^{I_0}_{\Delta\mu} &= 
\sum_{n',n}\sum_{\Delta\mu}\la\Phi_{n'}|\la\alpha,t|\Psi_{F_{\Delta\mu'+\Delta\mu}}\ra\la\Psi_{I_{\Delta\mu}}|\alpha,t\ra|\Phi_{n}\ra I^{F_0I_0}_{n',n}\\
&=\sum_{n',n}I^{F_0I_0}_{n',n}\sum_{\mu,\mu'}C^{F_{0}}_{n',\mu}C^{I_{0}^*}_{n,\mu'}\sum_{\Delta\mu}\la\alpha,t|N-\mu+\Delta\mu'+\Delta\mu\ra
\la N-\mu'+\Delta\mu|\alpha,t\ra,\nonumber
\end{align}
where
\begin{align}
I^{F_0I_0}_{n',n}=\sum_{\Delta\mu''}e^{i\Delta\mu''\omega t}\sum_{\mu}C^{F^*_{0}}_{n',\mu+\Delta\mu''} C^{I_0}_{n,\mu}.
\end{align}
Applying the approximation that $\sum_{\Delta\mu}A^*_{N-\mu+\Delta\mu'+\Delta\mu}A_{N-\mu'+\Delta\mu}\approx 1$ independently of $\mu'-\mu+\Delta\mu'$, we obtain that 

\begin{align}
\sum_{\Delta\mu}\la\alpha,t|N-\mu+\Delta\mu'+\Delta\mu\ra\la N-\mu'+\Delta\mu|\alpha,t\ra\approx e^{i(\mu'-\mu+\Delta\mu')\omega t}.
\end{align}
Thus, the sum $\sum_{\Delta\mu}B^{F_0^*}_{\Delta\mu'+\Delta\mu}B^{I_0}_{\Delta\mu} $ is given by $e^{i\Delta\mu'\omega t}\sum_{n',n} |I^{F_0I_0}_{n,n'}|^2$. It is further simplified using the following derivation
\begin{align}
\sum_{n',n} |I^{F_0I_0}_{n,n'}|^2 & = \sum_{\Delta\mu'}e^{i\Delta\mu'\omega t}\sum_{\mu,\mu'',n}C^{I_0}_{n,\mu}C^{I_0^*}_{n,\mu''+\Delta\mu'}\sum_{n',\Delta\mu}C^{F_0^*}_{n',\mu+\Delta\mu}C^{F_0}_{n',\mu''+\Delta\mu}\nonumber\\
& = \sum_{\Delta\mu',n}e^{i\Delta\mu'\omega t}\sum_{\mu,\mu''}C^{I_0}_{n,\mu}C^{I_0^*}_{n,\mu''+\Delta\mu'}\delta_{\mu,\mu''}\nonumber\\
& = \sum_{\Delta\mu'}e^{i\Delta\mu'\omega t}\sum_{\mu,n}C^{I_0}_{n,\mu}C^{I_0^*}_{n,\mu+\Delta\mu'}\nonumber\\
& = \sum_{\Delta\mu'}e^{i\Delta\mu'\omega t}\delta_{\Delta\mu',0}\nonumber\\
& = 1,
\end{align}
leading to a simple relation $\sum_{\Delta\mu}B^{F_0^*}_{\Delta\mu'+\Delta\mu}B^{I_0}_{\Delta\mu} = e^{i\Delta\mu'\omega t}$.
Therefore, we obtain that
\begin{align}
N_{F_0,I_0}(\br,t) = \sum_{\Delta\mu}e^{i\Delta\mu\omega t}\widetilde M_{F_0I_0}(\br,\Delta\mu),
\end{align}
resulting in the following relations for the matrix elements $\widetilde M_{F_0I_0}(\br,\Delta\mu)=\widetilde \rho_{F_0I_0}(\br,\Delta\mu)$:
\begin{align}
\widetilde M_{I_0I_0}(\br,\Delta\mu) & =  \sum_{i,\k}\sum_{m,m',\mu}c^{i*}_{m',\k,\mu+\Delta\mu}c^{i}_{m,\k,\mu}u_{m'\k}^\dagger(\br)u_{m\k}(\br)
\end{align}
and
\begin{align}
\widetilde M_{F_0\neq I_0I_0}(\br,\Delta\mu)
&= \sum_{m,m',\mu}c^{f*}_{m',\k',\mu+\Delta\mu}c^{i}_{m,\k,\mu}e^{i(\k-\k')\cdot\br}u_{m'\k'}^\dagger(\br)u_{m\k}(\br),
\end{align} 
where the same restrictions for the coefficients $\k$, $\k'$, $f$, $i$ apply as for Eqs.~(\ref{Eq_App_NI0I0}) and (\ref{Eq_App_NF0I0}), respectively.

\section{Evaluation of the Fourier transforms of functions $\widetilde M_{F_0I_0}(\br,\Delta\mu)$}
\label{App_Fourier}

\begin{widetext}
Let us consider Eq.~(\ref{Prob_General_coherent}), which depends on integrals $\int d^3 r e^{i\mathbf q_x\cdot\mathbf r}\widetilde M_{F_0I_0}(\br,\Delta\mu)$. Taking into account the periodicity of a crystal, we obtain that
\begin{align}
\int d^3 r e^{i\mathbf q_x\cdot\mathbf r}\widetilde M_{I_0I_0}(\br,\Delta\mu) 
&=  \sum_{i,\k} \sum_{m,m',\mu}c^{i*}_{m',\k,\mu+\Delta\mu}c^i_{m,\k,\mu}\int d^3r e^{i\mathbf q_x\cdot\mathbf r} u^\dagger_{m'\k}(\br)u_{m\k}(\br)\\
&=N_{\text{cells}}\sum_{i,\k}\sum_{m,m',\mu}c^{i*}_{m',\k,\mu+\Delta\mu}c^i_{m,\k,\mu} \sum_{\mathbf G}\delta(\mathbf q_x-\mathbf G) \int\limits_{V_{\text{cell}}} d^3r u_{m'\k}^\dagger(\br) u_{m\k}(\br)e^{i\mathbf G\cdot\mathbf r}\nonumber,
\end{align}
where $\delta(\mathbf q_x-\mathbf G)$ is the Dirac delta function, and
\begin{align}
\int d^3 r e^{i\mathbf q_x\cdot\mathbf r}\widetilde M_{F_0I_0\neq F_0}(\br,\Delta\mu) 
=& \sum_{m',m,\mu}c^{f*}_{m',\k',\mu+\Delta\mu}c^i_{m,\k\,mu}\int d^3 r e^{i(\mathbf q_x+\k-\k')\cdot\mathbf r}u^\dagger_{m'\k'}(\br)u_{m\k}(\br)\\
=&N_{\text{cells}} \sum_{m',m,\mu}c^{f*}_{m',\k',\mu+\Delta\mu}c^i_{m,\k,\mu}\sum_{\mathbf G} \delta(\mathbf q_x+\k-\k'-\mathbf G)\int\limits_{V_{\text{cell}}} d^3 ru^\dagger_{m'\k'}(\br)u_{m\k}(\br) e^{i\mathbf G\cdot\mathbf r},\nonumber
\end{align} 
where the same restrictions for the coefficients $\k$, $\k'$, $f$, $i$ apply as for Eqs.~(\ref{Eq_App_NI0I0}) and (\ref{Eq_App_NF0I0}), respectively.

Since we solve the time-dependent Schr\"odinger equation for the electronic system interacting with the pump pulse under the assumption that it can be solved separately for each $\k$ point, the number of electrons at each $\k$ point for all many-body states comprising the electronic wave function $|\Psi_\el,t\ra$ is the same. As a result, if states $K$ and $I$ comprise the wave function of the light-dressed system $|\Psi_0,t\ra$, $\int d^3 r e^{i\mathbf q_x\cdot\mathbf r}\widetilde M_{K_0I_0}(\br,\Delta\mu)$ are nonzero only for $\mathbf q_x = \mathbf G$.

\end{widetext}


\begin{thebibliography}{40}
\begin{thebibliography}{49}
\expandafter\ifx\csname natexlab\endcsname\relax\def\natexlab#1{#1}\fi
\expandafter\ifx\csname bibnamefont\endcsname\relax
  \def\bibnamefont#1{#1}\fi
\expandafter\ifx\csname bibfnamefont\endcsname\relax
  \def\bibfnamefont#1{#1}\fi
\expandafter\ifx\csname citenamefont\endcsname\relax
  \def\citenamefont#1{#1}\fi
\expandafter\ifx\csname url\endcsname\relax
  \def\url#1{\texttt{#1}}\fi
\expandafter\ifx\csname urlprefix\endcsname\relax\def\urlprefix{URL }\fi
\providecommand{\bibinfo}[2]{#2}
\providecommand{\eprint}[2][]{\url{#2}}

\bibitem[{\citenamefont{Shirley}(1965)}]{ShirleyPR65}
\bibinfo{author}{\bibfnamefont{J.~H.} \bibnamefont{Shirley}},
  \bibinfo{journal}{Phys. Rev.} \textbf{\bibinfo{volume}{138}},
  \bibinfo{pages}{B979} (\bibinfo{year}{1965}).

\bibitem[{\citenamefont{Santra and Greene}(2004)}]{SantraPRA04}
\bibinfo{author}{\bibfnamefont{R.}~\bibnamefont{Santra}} \bibnamefont{and}
  \bibinfo{author}{\bibfnamefont{C.~H.} \bibnamefont{Greene}},
  \bibinfo{journal}{Phys. Rev. A} \textbf{\bibinfo{volume}{70}},
  \bibinfo{pages}{053401} (\bibinfo{year}{2004}).

\bibitem[{\citenamefont{Oka and Aoki}(2009)}]{OkaPRB09}
\bibinfo{author}{\bibfnamefont{T.}~\bibnamefont{Oka}} \bibnamefont{and}
  \bibinfo{author}{\bibfnamefont{H.}~\bibnamefont{Aoki}},
  \bibinfo{journal}{Phys. Rev. B} \textbf{\bibinfo{volume}{79}},
  \bibinfo{pages}{081406} (\bibinfo{year}{2009}).

\bibitem[{\citenamefont{Struck et~al.}(2011)\citenamefont{Struck,
  {\"O}lschl{\"a}ger, Le~Targat, Soltan-Panahi, Eckardt, Lewenstein,
  Windpassinger, and Sengstock}}]{StruckScience11}
\bibinfo{author}{\bibfnamefont{J.}~\bibnamefont{Struck}},
  \bibinfo{author}{\bibfnamefont{C.}~\bibnamefont{{\"O}lschl{\"a}ger}},
  \bibinfo{author}{\bibfnamefont{R.}~\bibnamefont{Le~Targat}},
  \bibinfo{author}{\bibfnamefont{P.}~\bibnamefont{Soltan-Panahi}},
  \bibinfo{author}{\bibfnamefont{A.}~\bibnamefont{Eckardt}},
  \bibinfo{author}{\bibfnamefont{M.}~\bibnamefont{Lewenstein}},
  \bibinfo{author}{\bibfnamefont{P.}~\bibnamefont{Windpassinger}},
  \bibnamefont{and}
  \bibinfo{author}{\bibfnamefont{K.}~\bibnamefont{Sengstock}},
  \bibinfo{journal}{Science} \textbf{\bibinfo{volume}{333}},
  \bibinfo{pages}{996} (\bibinfo{year}{2011}).

\bibitem[{\citenamefont{Aidelsburger et~al.}(2011)\citenamefont{Aidelsburger,
  Atala, Nascimb\`ene, Trotzky, Chen, and Bloch}}]{AidelsburgerPRL11}
\bibinfo{author}{\bibfnamefont{M.}~\bibnamefont{Aidelsburger}},
  \bibinfo{author}{\bibfnamefont{M.}~\bibnamefont{Atala}},
  \bibinfo{author}{\bibfnamefont{S.}~\bibnamefont{Nascimb\`ene}},
  \bibinfo{author}{\bibfnamefont{S.}~\bibnamefont{Trotzky}},
  \bibinfo{author}{\bibfnamefont{Y.-A.} \bibnamefont{Chen}}, \bibnamefont{and}
  \bibinfo{author}{\bibfnamefont{I.}~\bibnamefont{Bloch}},
  \bibinfo{journal}{Phys. Rev. Lett.} \textbf{\bibinfo{volume}{107}},
  \bibinfo{pages}{255301} (\bibinfo{year}{2011}).

\bibitem[{\citenamefont{Wang et~al.}(2013)\citenamefont{Wang, Steinberg,
  Jarillo-Herrero, and Gedik}}]{WangScience13}
\bibinfo{author}{\bibfnamefont{Y.~H.} \bibnamefont{Wang}},
  \bibinfo{author}{\bibfnamefont{H.}~\bibnamefont{Steinberg}},
  \bibinfo{author}{\bibfnamefont{P.}~\bibnamefont{Jarillo-Herrero}},
  \bibnamefont{and} \bibinfo{author}{\bibfnamefont{N.}~\bibnamefont{Gedik}},
  \bibinfo{journal}{Science} \textbf{\bibinfo{volume}{342}},
  \bibinfo{pages}{453} (\bibinfo{year}{2013}).

\bibitem[{\citenamefont{Goldman and Dalibard}(2014)}]{GoldmanPRX14}
\bibinfo{author}{\bibfnamefont{N.}~\bibnamefont{Goldman}} \bibnamefont{and}
  \bibinfo{author}{\bibfnamefont{J.}~\bibnamefont{Dalibard}},
  \bibinfo{journal}{Phys. Rev. X} \textbf{\bibinfo{volume}{4}},
  \bibinfo{pages}{031027} (\bibinfo{year}{2014}).

\bibitem[{\citenamefont{H\"ubener et~al.}(2017)\citenamefont{H\"ubener, Sentef,
  De~Giovannini, Kemper, and Rubio}}]{HuebenerNature17}
\bibinfo{author}{\bibfnamefont{H.}~\bibnamefont{H\"ubener}},
  \bibinfo{author}{\bibfnamefont{M.~A.} \bibnamefont{Sentef}},
  \bibinfo{author}{\bibfnamefont{U.}~\bibnamefont{De~Giovannini}},
  \bibinfo{author}{\bibfnamefont{A.~F.} \bibnamefont{Kemper}},
  \bibnamefont{and} \bibinfo{author}{\bibfnamefont{A.}~\bibnamefont{Rubio}},
  \bibinfo{journal}{Nature Communications} \textbf{\bibinfo{volume}{8}},
  \bibinfo{pages}{13940} (\bibinfo{year}{2017}).

\bibitem[{\citenamefont{Oka and Kitamura}(2018)}]{OkaArxiv18}
\bibinfo{author}{\bibfnamefont{T.}~\bibnamefont{Oka}} \bibnamefont{and}
  \bibinfo{author}{\bibfnamefont{S.}~\bibnamefont{Kitamura}},
  \bibinfo{journal}{arXiv:1804.03212}  (\bibinfo{year}{2018}).

\bibitem[{\citenamefont{Faisal and Kami\ifmmode~\acute{n}\else
  \'{n}\fi{}ski}(1997)}]{FaisalPRA97}
\bibinfo{author}{\bibfnamefont{F.~H.~M.} \bibnamefont{Faisal}}
  \bibnamefont{and} \bibinfo{author}{\bibfnamefont{J.~Z.}
  \bibnamefont{Kami\ifmmode~\acute{n}\else \'{n}\fi{}ski}},
  \bibinfo{journal}{Phys. Rev. A} \textbf{\bibinfo{volume}{56}},
  \bibinfo{pages}{748} (\bibinfo{year}{1997}).

\bibitem[{\citenamefont{Chu and Telnov}(2004)}]{ChuPhRep04}
\bibinfo{author}{\bibfnamefont{S.-I.} \bibnamefont{Chu}} \bibnamefont{and}
  \bibinfo{author}{\bibfnamefont{D.~A.} \bibnamefont{Telnov}},
  \bibinfo{journal}{Physics Reports} \textbf{\bibinfo{volume}{390}},
  \bibinfo{pages}{1} (\bibinfo{year}{2004}).

\bibitem[{\citenamefont{Buth et~al.}(2007)\citenamefont{Buth, Santra, and
  Young}}]{ButhPRL07}
\bibinfo{author}{\bibfnamefont{C.}~\bibnamefont{Buth}},
  \bibinfo{author}{\bibfnamefont{R.}~\bibnamefont{Santra}}, \bibnamefont{and}
  \bibinfo{author}{\bibfnamefont{L.}~\bibnamefont{Young}},
  \bibinfo{journal}{Phys. Rev. Lett.} \textbf{\bibinfo{volume}{98}},
  \bibinfo{pages}{253001} (\bibinfo{year}{2007}).

\bibitem[{\citenamefont{Smirnova et~al.}(2009)\citenamefont{Smirnova, Mairesse,
  Patchkovskii, Dudovich, Villeneuve, Corkum, and Ivanov}}]{SmirnovaNature09}
\bibinfo{author}{\bibfnamefont{O.}~\bibnamefont{Smirnova}},
  \bibinfo{author}{\bibfnamefont{Y.}~\bibnamefont{Mairesse}},
  \bibinfo{author}{\bibfnamefont{S.}~\bibnamefont{Patchkovskii}},
  \bibinfo{author}{\bibfnamefont{N.}~\bibnamefont{Dudovich}},
  \bibinfo{author}{\bibfnamefont{D.}~\bibnamefont{Villeneuve}},
  \bibinfo{author}{\bibfnamefont{P.}~\bibnamefont{Corkum}}, \bibnamefont{and}
  \bibinfo{author}{\bibfnamefont{M.~Y.} \bibnamefont{Ivanov}},
  \bibinfo{journal}{Nature} \textbf{\bibinfo{volume}{460}},
  \bibinfo{pages}{972} (\bibinfo{year}{2009}).

\bibitem[{\citenamefont{Ghimire et~al.}(2010)\citenamefont{Ghimire, DiChiara,
  Sistrunk, Agostini, DiMauro, and Reis}}]{GhimireNature11}
\bibinfo{author}{\bibfnamefont{S.}~\bibnamefont{Ghimire}},
  \bibinfo{author}{\bibfnamefont{A.~D.} \bibnamefont{DiChiara}},
  \bibinfo{author}{\bibfnamefont{E.}~\bibnamefont{Sistrunk}},
  \bibinfo{author}{\bibfnamefont{P.}~\bibnamefont{Agostini}},
  \bibinfo{author}{\bibfnamefont{L.~F.} \bibnamefont{DiMauro}},
  \bibnamefont{and} \bibinfo{author}{\bibfnamefont{D.~A.} \bibnamefont{Reis}},
  \bibinfo{journal}{Nature Physics} \textbf{\bibinfo{volume}{7}},
  \bibinfo{pages}{138} (\bibinfo{year}{2010}).

\bibitem[{\citenamefont{Schubert et~al.}(2014)\citenamefont{Schubert,
  Hohenleutner, Langer, Urbanek, Lange, Huttner, Golde, Meier, Kira, Koch
  et~al.}}]{SchubertNature14}
\bibinfo{author}{\bibfnamefont{O.}~\bibnamefont{Schubert}},
  \bibinfo{author}{\bibfnamefont{M.}~\bibnamefont{Hohenleutner}},
  \bibinfo{author}{\bibfnamefont{F.}~\bibnamefont{Langer}},
  \bibinfo{author}{\bibfnamefont{B.}~\bibnamefont{Urbanek}},
  \bibinfo{author}{\bibfnamefont{C.}~\bibnamefont{Lange}},
  \bibinfo{author}{\bibfnamefont{U.}~\bibnamefont{Huttner}},
  \bibinfo{author}{\bibfnamefont{D.}~\bibnamefont{Golde}},
  \bibinfo{author}{\bibfnamefont{T.}~\bibnamefont{Meier}},
  \bibinfo{author}{\bibfnamefont{M.}~\bibnamefont{Kira}},
  \bibinfo{author}{\bibfnamefont{S.~W.} \bibnamefont{Koch}},
  \bibnamefont{et~al.}, \bibinfo{journal}{Nature Photonics}
  \textbf{\bibinfo{volume}{8}}, \bibinfo{pages}{119} (\bibinfo{year}{2014}).

\bibitem[{\citenamefont{Corkum and Krausz}(2007)}]{CorkumNature07}
\bibinfo{author}{\bibfnamefont{P.}~\bibnamefont{Corkum}} \bibnamefont{and}
  \bibinfo{author}{\bibfnamefont{F.}~\bibnamefont{Krausz}},
  \bibinfo{journal}{Nature Physics} \textbf{\bibinfo{volume}{3}},
  \bibinfo{pages}{381} (\bibinfo{year}{2007}).

\bibitem[{\citenamefont{Gaffney and Chapman}(2007)}]{GaffneyScience07}
\bibinfo{author}{\bibfnamefont{K.~J.} \bibnamefont{Gaffney}} \bibnamefont{and}
  \bibinfo{author}{\bibfnamefont{H.~N.} \bibnamefont{Chapman}},
  \bibinfo{journal}{Science} \textbf{\bibinfo{volume}{316}},
  \bibinfo{pages}{1444} (\bibinfo{year}{2007}).

\bibitem[{\citenamefont{Chapman et~al.}(2006)\citenamefont{Chapman, Barty,
  Bogan, Boutet, Frank, Hau-Riege, Marchesini, Woods, Bajt, Benner
  et~al.}}]{ChapmanNature06}
\bibinfo{author}{\bibfnamefont{H.~N.} \bibnamefont{Chapman}},
  \bibinfo{author}{\bibfnamefont{A.}~\bibnamefont{Barty}},
  \bibinfo{author}{\bibfnamefont{M.~J.} \bibnamefont{Bogan}},
  \bibinfo{author}{\bibfnamefont{S.}~\bibnamefont{Boutet}},
  \bibinfo{author}{\bibfnamefont{M.}~\bibnamefont{Frank}},
  \bibinfo{author}{\bibfnamefont{S.~P.} \bibnamefont{Hau-Riege}},
  \bibinfo{author}{\bibfnamefont{S.}~\bibnamefont{Marchesini}},
  \bibinfo{author}{\bibfnamefont{B.~W.} \bibnamefont{Woods}},
  \bibinfo{author}{\bibfnamefont{S.}~\bibnamefont{Bajt}},
  \bibinfo{author}{\bibfnamefont{H.}~\bibnamefont{Benner}},
  \bibnamefont{et~al.}, \bibinfo{journal}{Nature Physics}
  \textbf{\bibinfo{volume}{2}}, \bibinfo{pages}{839} (\bibinfo{year}{2006}).

\bibitem[{\citenamefont{Vrakking and Elsaesser}(2012)}]{VrakkingNature12}
\bibinfo{author}{\bibfnamefont{M.~J.~J.} \bibnamefont{Vrakking}}
  \bibnamefont{and}
  \bibinfo{author}{\bibfnamefont{T.}~\bibnamefont{Elsaesser}},
  \bibinfo{journal}{Nature Photonics} \textbf{\bibinfo{volume}{6}},
  \bibinfo{pages}{645} (\bibinfo{year}{2012}).

\bibitem[{\citenamefont{Leone et~al.}(2014)\citenamefont{Leone, McCurdy,
  Burgdorfer, Cederbaum, Chang, Dudovich, Feist, Greene, Ivanov, Kienberger
  et~al.}}]{LeoneNature14}
\bibinfo{author}{\bibfnamefont{S.~R.} \bibnamefont{Leone}},
  \bibinfo{author}{\bibfnamefont{C.~W.} \bibnamefont{McCurdy}},
  \bibinfo{author}{\bibfnamefont{J.}~\bibnamefont{Burgdorfer}},
  \bibinfo{author}{\bibfnamefont{L.~S.} \bibnamefont{Cederbaum}},
  \bibinfo{author}{\bibfnamefont{Z.}~\bibnamefont{Chang}},
  \bibinfo{author}{\bibfnamefont{N.}~\bibnamefont{Dudovich}},
  \bibinfo{author}{\bibfnamefont{J.}~\bibnamefont{Feist}},
  \bibinfo{author}{\bibfnamefont{C.~H.} \bibnamefont{Greene}},
  \bibinfo{author}{\bibfnamefont{M.}~\bibnamefont{Ivanov}},
  \bibinfo{author}{\bibfnamefont{R.}~\bibnamefont{Kienberger}},
  \bibnamefont{et~al.}, \bibinfo{journal}{Nature Photonics}
  \textbf{\bibinfo{volume}{8}}, \bibinfo{pages}{162} (\bibinfo{year}{2014}).

\bibitem[{\citenamefont{Kowalewski et~al.}(2017)\citenamefont{Kowalewski,
  Fingerhut, Dorfman, Bennett, and Mukamel}}]{KowalewskiChRev17}
\bibinfo{author}{\bibfnamefont{M.}~\bibnamefont{Kowalewski}},
  \bibinfo{author}{\bibfnamefont{B.~P.} \bibnamefont{Fingerhut}},
  \bibinfo{author}{\bibfnamefont{K.~E.} \bibnamefont{Dorfman}},
  \bibinfo{author}{\bibfnamefont{K.}~\bibnamefont{Bennett}}, \bibnamefont{and}
  \bibinfo{author}{\bibfnamefont{S.}~\bibnamefont{Mukamel}},
  \bibinfo{journal}{Chemical Reviews} \textbf{\bibinfo{volume}{117}},
  \bibinfo{pages}{12165} (\bibinfo{year}{2017}), \bibinfo{note}{pMID:
  28949133}.

\bibitem[{\citenamefont{Neville et~al.}(2018)\citenamefont{Neville, Chergui,
  Stolow, and Schuurman}}]{NevillePRL18}
\bibinfo{author}{\bibfnamefont{S.~P.} \bibnamefont{Neville}},
  \bibinfo{author}{\bibfnamefont{M.}~\bibnamefont{Chergui}},
  \bibinfo{author}{\bibfnamefont{A.}~\bibnamefont{Stolow}}, \bibnamefont{and}
  \bibinfo{author}{\bibfnamefont{M.~S.} \bibnamefont{Schuurman}},
  \bibinfo{journal}{Phys. Rev. Lett.} \textbf{\bibinfo{volume}{120}},
  \bibinfo{pages}{243001} (\bibinfo{year}{2018}).

\bibitem[{\citenamefont{Dixit et~al.}(2012)\citenamefont{Dixit, Vendrell, and
  Santra}}]{DixitPNAS12}
\bibinfo{author}{\bibfnamefont{G.}~\bibnamefont{Dixit}},
  \bibinfo{author}{\bibfnamefont{O.}~\bibnamefont{Vendrell}}, \bibnamefont{and}
  \bibinfo{author}{\bibfnamefont{R.}~\bibnamefont{Santra}},
  \bibinfo{journal}{Proceedings of the National Academy of Sciences}
  \textbf{\bibinfo{volume}{109}}, \bibinfo{pages}{11636}
  (\bibinfo{year}{2012}).

\bibitem[{\citenamefont{Popova-Gorelova and
  Santra}(2015{\natexlab{a}})}]{PopovaGorelovaPRB15_1}
\bibinfo{author}{\bibfnamefont{D.}~\bibnamefont{Popova-Gorelova}}
  \bibnamefont{and} \bibinfo{author}{\bibfnamefont{R.}~\bibnamefont{Santra}},
  \bibinfo{journal}{Phys. Rev. B} \textbf{\bibinfo{volume}{91}},
  \bibinfo{pages}{184303} (\bibinfo{year}{2015}{\natexlab{a}}).

\bibitem[{\citenamefont{Popova-Gorelova and
  Santra}(2015{\natexlab{b}})}]{PopovaGorelovaPRB15_2}
\bibinfo{author}{\bibfnamefont{D.}~\bibnamefont{Popova-Gorelova}}
  \bibnamefont{and} \bibinfo{author}{\bibfnamefont{R.}~\bibnamefont{Santra}},
  \bibinfo{journal}{Phys. Rev. B} \textbf{\bibinfo{volume}{92}},
  \bibinfo{pages}{184304} (\bibinfo{year}{2015}{\natexlab{b}}).

\bibitem[{\citenamefont{Mandel and Wolf}(1995)}]{Mandel}
\bibinfo{author}{\bibfnamefont{L.}~\bibnamefont{Mandel}} \bibnamefont{and}
  \bibinfo{author}{\bibfnamefont{E.}~\bibnamefont{Wolf}},
  \emph{\bibinfo{title}{Optical Coherence and Quantum Optics}}
  (\bibinfo{publisher}{Cambridge University Press},
  \bibinfo{address}{Cambridge}, \bibinfo{year}{1995}).

\bibitem[{\citenamefont{Popova-Gorelova}(2018)}]{Popova-GorelovaAppSci18}
\bibinfo{author}{\bibfnamefont{D.}~\bibnamefont{Popova-Gorelova}},
  \bibinfo{journal}{Applied Sciences} \textbf{\bibinfo{volume}{8}},
  \bibinfo{pages}{318} (\bibinfo{year}{2018}).

\bibitem[{\citenamefont{Freund and Levine}(1970)}]{FreundPRL70}
\bibinfo{author}{\bibfnamefont{I.}~\bibnamefont{Freund}} \bibnamefont{and}
  \bibinfo{author}{\bibfnamefont{B.~F.} \bibnamefont{Levine}},
  \bibinfo{journal}{Phys. Rev. Lett.} \textbf{\bibinfo{volume}{25}},
  \bibinfo{pages}{1241} (\bibinfo{year}{1970}).

\bibitem[{\citenamefont{Eisenberger and McCall}(1971)}]{EisenbergerPRA71}
\bibinfo{author}{\bibfnamefont{P.~M.} \bibnamefont{Eisenberger}}
  \bibnamefont{and} \bibinfo{author}{\bibfnamefont{S.~L.}
  \bibnamefont{McCall}}, \bibinfo{journal}{Phys. Rev. A}
  \textbf{\bibinfo{volume}{3}}, \bibinfo{pages}{1145} (\bibinfo{year}{1971}).

\bibitem[{\citenamefont{Glover et~al.}(2012)\citenamefont{Glover, Fritz,
  Cammarata, Allison, Coh, Feldkamp, Lemke, Zhu, Feng, Coffee
  et~al.}}]{GloverNature12}
\bibinfo{author}{\bibfnamefont{T.~E.} \bibnamefont{Glover}},
  \bibinfo{author}{\bibfnamefont{D.~M.} \bibnamefont{Fritz}},
  \bibinfo{author}{\bibfnamefont{M.}~\bibnamefont{Cammarata}},
  \bibinfo{author}{\bibfnamefont{T.~K.} \bibnamefont{Allison}},
  \bibinfo{author}{\bibfnamefont{S.}~\bibnamefont{Coh}},
  \bibinfo{author}{\bibfnamefont{J.~M.} \bibnamefont{Feldkamp}},
  \bibinfo{author}{\bibfnamefont{H.}~\bibnamefont{Lemke}},
  \bibinfo{author}{\bibfnamefont{D.}~\bibnamefont{Zhu}},
  \bibinfo{author}{\bibfnamefont{Y.}~\bibnamefont{Feng}},
  \bibinfo{author}{\bibfnamefont{R.~N.} \bibnamefont{Coffee}},
  \bibnamefont{et~al.}, \bibinfo{journal}{Nature}
  \textbf{\bibinfo{volume}{488}}, \bibinfo{pages}{603} (\bibinfo{year}{2012}).

\bibitem[{\citenamefont{Krausz and Ivanov}(2009)}]{KrauszRMP09}
\bibinfo{author}{\bibfnamefont{F.}~\bibnamefont{Krausz}} \bibnamefont{and}
  \bibinfo{author}{\bibfnamefont{M.}~\bibnamefont{Ivanov}},
  \bibinfo{journal}{Rev. Mod. Phys.} \textbf{\bibinfo{volume}{81}},
  \bibinfo{pages}{163} (\bibinfo{year}{2009}).

\bibitem[{\citenamefont{Vampa et~al.}(2015)\citenamefont{Vampa, McDonald,
  Orlando, Corkum, and Brabec}}]{VampaPRB15}
\bibinfo{author}{\bibfnamefont{G.}~\bibnamefont{Vampa}},
  \bibinfo{author}{\bibfnamefont{C.~R.} \bibnamefont{McDonald}},
  \bibinfo{author}{\bibfnamefont{G.}~\bibnamefont{Orlando}},
  \bibinfo{author}{\bibfnamefont{P.~B.} \bibnamefont{Corkum}},
  \bibnamefont{and} \bibinfo{author}{\bibfnamefont{T.}~\bibnamefont{Brabec}},
  \bibinfo{journal}{Phys. Rev. B} \textbf{\bibinfo{volume}{91}},
  \bibinfo{pages}{064302} (\bibinfo{year}{2015}).

\bibitem[{\citenamefont{McDonald et~al.}(2015)\citenamefont{McDonald, Vampa,
  Corkum, and Brabec}}]{McDonaldPRA15}
\bibinfo{author}{\bibfnamefont{C.~R.} \bibnamefont{McDonald}},
  \bibinfo{author}{\bibfnamefont{G.}~\bibnamefont{Vampa}},
  \bibinfo{author}{\bibfnamefont{P.~B.} \bibnamefont{Corkum}},
  \bibnamefont{and} \bibinfo{author}{\bibfnamefont{T.}~\bibnamefont{Brabec}},
  \bibinfo{journal}{Phys. Rev. A} \textbf{\bibinfo{volume}{92}},
  \bibinfo{pages}{033845} (\bibinfo{year}{2015}).

\bibitem[{\citenamefont{Ndabashimiye et~al.}(2016)\citenamefont{Ndabashimiye,
  Ghimire, Wu, Browne, Schafer, Gaarde, and Reis}}]{NdabashimiyeNature16}
\bibinfo{author}{\bibfnamefont{G.}~\bibnamefont{Ndabashimiye}},
  \bibinfo{author}{\bibfnamefont{S.}~\bibnamefont{Ghimire}},
  \bibinfo{author}{\bibfnamefont{M.}~\bibnamefont{Wu}},
  \bibinfo{author}{\bibfnamefont{D.~A.} \bibnamefont{Browne}},
  \bibinfo{author}{\bibfnamefont{K.~J.} \bibnamefont{Schafer}},
  \bibinfo{author}{\bibfnamefont{M.~B.} \bibnamefont{Gaarde}},
  \bibnamefont{and} \bibinfo{author}{\bibfnamefont{D.~A.} \bibnamefont{Reis}},
  \bibinfo{journal}{Nature} \textbf{\bibinfo{volume}{534}},
  \bibinfo{pages}{520} (\bibinfo{year}{2016}).

\bibitem[{\citenamefont{Tamaya et~al.}(2016)\citenamefont{Tamaya, Ishikawa,
  Ogawa, and Tanaka}}]{TamayaPRL16}
\bibinfo{author}{\bibfnamefont{T.}~\bibnamefont{Tamaya}},
  \bibinfo{author}{\bibfnamefont{A.}~\bibnamefont{Ishikawa}},
  \bibinfo{author}{\bibfnamefont{T.}~\bibnamefont{Ogawa}}, \bibnamefont{and}
  \bibinfo{author}{\bibfnamefont{K.}~\bibnamefont{Tanaka}},
  \bibinfo{journal}{Phys. Rev. Lett.} \textbf{\bibinfo{volume}{116}},
  \bibinfo{pages}{016601} (\bibinfo{year}{2016}).

\bibitem[{\citenamefont{Luu and W\"orner}(2016)}]{LuuPRB16}
\bibinfo{author}{\bibfnamefont{T.~T.} \bibnamefont{Luu}} \bibnamefont{and}
  \bibinfo{author}{\bibfnamefont{H.~J.} \bibnamefont{W\"orner}},
  \bibinfo{journal}{Phys. Rev. B} \textbf{\bibinfo{volume}{94}},
  \bibinfo{pages}{115164} (\bibinfo{year}{2016}).

\bibitem[{\citenamefont{Tancogne-Dejean
  et~al.}(2017)\citenamefont{Tancogne-Dejean, M\"ucke, K\"artner, and
  Rubio}}]{Tancogne-DejeanPRL17}
\bibinfo{author}{\bibfnamefont{N.}~\bibnamefont{Tancogne-Dejean}},
  \bibinfo{author}{\bibfnamefont{O.~D.} \bibnamefont{M\"ucke}},
  \bibinfo{author}{\bibfnamefont{F.~X.} \bibnamefont{K\"artner}},
  \bibnamefont{and} \bibinfo{author}{\bibfnamefont{A.}~\bibnamefont{Rubio}},
  \bibinfo{journal}{Phys. Rev. Lett.} \textbf{\bibinfo{volume}{118}},
  \bibinfo{pages}{087403} (\bibinfo{year}{2017}).

\bibitem[{\citenamefont{Ghimire et~al.}(2014)\citenamefont{Ghimire,
  Ndabashimiye, DiChiara, Sistrunk, Stockman, Agostini, DiMauro, and
  Reis}}]{GhimireJPhB14}
\bibinfo{author}{\bibfnamefont{S.}~\bibnamefont{Ghimire}},
  \bibinfo{author}{\bibfnamefont{G.}~\bibnamefont{Ndabashimiye}},
  \bibinfo{author}{\bibfnamefont{A.~D.} \bibnamefont{DiChiara}},
  \bibinfo{author}{\bibfnamefont{E.}~\bibnamefont{Sistrunk}},
  \bibinfo{author}{\bibfnamefont{M.~I.} \bibnamefont{Stockman}},
  \bibinfo{author}{\bibfnamefont{P.}~\bibnamefont{Agostini}},
  \bibinfo{author}{\bibfnamefont{L.~F.} \bibnamefont{DiMauro}},
  \bibnamefont{and} \bibinfo{author}{\bibfnamefont{D.~A.} \bibnamefont{Reis}},
  \bibinfo{journal}{Journal of Physics B: Atomic, Molecular and Optical
  Physics} \textbf{\bibinfo{volume}{47}}, \bibinfo{pages}{204030}
  (\bibinfo{year}{2014}).

\bibitem[{\citenamefont{You et~al.}(2016)\citenamefont{You, Reis, and
  Ghimire}}]{YouNature16}
\bibinfo{author}{\bibfnamefont{Y.~S.} \bibnamefont{You}},
  \bibinfo{author}{\bibfnamefont{D.~A.} \bibnamefont{Reis}}, \bibnamefont{and}
  \bibinfo{author}{\bibfnamefont{S.}~\bibnamefont{Ghimire}},
  \bibinfo{journal}{Nature Physics} \textbf{\bibinfo{volume}{13}},
  \bibinfo{pages}{345} (\bibinfo{year}{2016}).

\bibitem[{\citenamefont{Drake}(2006)}]{SpringerBookAMO}
\bibinfo{author}{\bibfnamefont{G.}~\bibnamefont{Drake}},
  \emph{\bibinfo{title}{Springer Handbook of Atomic, Molecular, and Optical
  Physics}} (\bibinfo{publisher}{Springer}, \bibinfo{year}{2006}).

\bibitem[{\citenamefont{Hsu and Reichl}(2006)}]{HsuPRB06}
\bibinfo{author}{\bibfnamefont{H.}~\bibnamefont{Hsu}} \bibnamefont{and}
  \bibinfo{author}{\bibfnamefont{L.~E.} \bibnamefont{Reichl}},
  \bibinfo{journal}{Phys. Rev. B} \textbf{\bibinfo{volume}{74}},
  \bibinfo{pages}{115406} (\bibinfo{year}{2006}).

\bibitem[{\citenamefont{Glauber}(1963)}]{GlauberPhRev63}
\bibinfo{author}{\bibfnamefont{R.~J.} \bibnamefont{Glauber}},
  \bibinfo{journal}{Phys. Rev.} \textbf{\bibinfo{volume}{130}},
  \bibinfo{pages}{2529} (\bibinfo{year}{1963}).

\bibitem[{\citenamefont{Loudon}(1983)}]{Loudon}
\bibinfo{author}{\bibfnamefont{R.}~\bibnamefont{Loudon}},
  \emph{\bibinfo{title}{The quantum theory of light}}
  (\bibinfo{publisher}{Clarendon Press}, \bibinfo{address}{Oxford},
  \bibinfo{year}{1983}).

\bibitem[{\citenamefont{Kittel}(2004)}]{Kittel}
\bibinfo{author}{\bibfnamefont{C.}~\bibnamefont{Kittel}},
  \emph{\bibinfo{title}{Introduction to Solid State Physics}}
  (\bibinfo{publisher}{Wiley}, \bibinfo{address}{New York City},
  \bibinfo{year}{2004}).

\bibitem[{\citenamefont{Tzoar and Gersten}(1975)}]{TzoarPRB75}
\bibinfo{author}{\bibfnamefont{N.}~\bibnamefont{Tzoar}} \bibnamefont{and}
  \bibinfo{author}{\bibfnamefont{J.~I.} \bibnamefont{Gersten}},
  \bibinfo{journal}{Phys. Rev. B} \textbf{\bibinfo{volume}{12}},
  \bibinfo{pages}{1132} (\bibinfo{year}{1975}).

\bibitem[{\citenamefont{Gonze et~al.}(2016)\citenamefont{Gonze, Jollet, Araujo,
  Adams, Amadon, Applencourt, Audouze, Beuken, Bieder, Bokhanchuk
  et~al.}}]{Gonze16}
\bibinfo{author}{\bibfnamefont{X.}~\bibnamefont{Gonze}},
  \bibinfo{author}{\bibfnamefont{F.}~\bibnamefont{Jollet}},
  \bibinfo{author}{\bibfnamefont{F.~A.} \bibnamefont{Araujo}},
  \bibinfo{author}{\bibfnamefont{D.}~\bibnamefont{Adams}},
  \bibinfo{author}{\bibfnamefont{B.}~\bibnamefont{Amadon}},
  \bibinfo{author}{\bibfnamefont{T.}~\bibnamefont{Applencourt}},
  \bibinfo{author}{\bibfnamefont{C.}~\bibnamefont{Audouze}},
  \bibinfo{author}{\bibfnamefont{J.-M.} \bibnamefont{Beuken}},
  \bibinfo{author}{\bibfnamefont{J.}~\bibnamefont{Bieder}},
  \bibinfo{author}{\bibfnamefont{A.}~\bibnamefont{Bokhanchuk}},
  \bibnamefont{et~al.}, \bibinfo{journal}{Computer Physics Communications}
  \textbf{\bibinfo{volume}{205}}, \bibinfo{pages}{106 } (\bibinfo{year}{2016}).

\bibitem[{\citenamefont{Gonze et~al.}(2009)\citenamefont{Gonze, Amadon,
  Anglade, Beuken, Bottin, Boulanger, Bruneval, Caliste, Caracas, CÃŽtÃ©
  et~al.}}]{Gonze09}
\bibinfo{author}{\bibfnamefont{X.}~\bibnamefont{Gonze}},
  \bibinfo{author}{\bibfnamefont{B.}~\bibnamefont{Amadon}},
  \bibinfo{author}{\bibfnamefont{P.-M.} \bibnamefont{Anglade}},
  \bibinfo{author}{\bibfnamefont{J.-M.} \bibnamefont{Beuken}},
  \bibinfo{author}{\bibfnamefont{F.}~\bibnamefont{Bottin}},
  \bibinfo{author}{\bibfnamefont{P.}~\bibnamefont{Boulanger}},
  \bibinfo{author}{\bibfnamefont{F.}~\bibnamefont{Bruneval}},
  \bibinfo{author}{\bibfnamefont{D.}~\bibnamefont{Caliste}},
  \bibinfo{author}{\bibfnamefont{R.}~\bibnamefont{Caracas}},
  \bibinfo{author}{\bibfnamefont{M.}~\bibnamefont{CÃŽtÃ©}},
  \bibnamefont{et~al.}, \bibinfo{journal}{Computer Physics Communications}
  \textbf{\bibinfo{volume}{180}}, \bibinfo{pages}{2582 }
  (\bibinfo{year}{2009}).

\bibitem[{\citenamefont{Gonze et~al.}(2005)\citenamefont{Gonze, Rignanese,
  Verstraete, Beuken, Pouillon, Caracas, Jollet, Torrent, Zerah, Mikami
  et~al.}}]{Gonze05}
\bibinfo{author}{\bibfnamefont{X.}~\bibnamefont{Gonze}},
  \bibinfo{author}{\bibfnamefont{G.}~\bibnamefont{Rignanese}},
  \bibinfo{author}{\bibfnamefont{M.}~\bibnamefont{Verstraete}},
  \bibinfo{author}{\bibfnamefont{J.}~\bibnamefont{Beuken}},
  \bibinfo{author}{\bibfnamefont{Y.}~\bibnamefont{Pouillon}},
  \bibinfo{author}{\bibfnamefont{R.}~\bibnamefont{Caracas}},
  \bibinfo{author}{\bibfnamefont{F.}~\bibnamefont{Jollet}},
  \bibinfo{author}{\bibfnamefont{M.}~\bibnamefont{Torrent}},
  \bibinfo{author}{\bibfnamefont{G.}~\bibnamefont{Zerah}},
  \bibinfo{author}{\bibfnamefont{M.}~\bibnamefont{Mikami}},
  \bibnamefont{et~al.}, \bibinfo{journal}{Zeitschrift f{\"u}r Kristallographie}
  \textbf{\bibinfo{volume}{220}}, \bibinfo{pages}{558} (\bibinfo{year}{2005}).

\bibitem[{\citenamefont{Troullier and
  Martins}(1991)}]{Troullier-Martins_Pseudpotentials}
\bibinfo{author}{\bibfnamefont{N.}~\bibnamefont{Troullier}} \bibnamefont{and}
  \bibinfo{author}{\bibfnamefont{J.~L.} \bibnamefont{Martins}},
  \bibinfo{journal}{Phys. Rev. B} \textbf{\bibinfo{volume}{43}},
  \bibinfo{pages}{1993} (\bibinfo{year}{1991}).

\end{thebibliography}

\end{thebibliography}

\end{document}